\documentclass[useAMS,usenatbib]{mn2e}
\usepackage{amsmath}
\usepackage{url}
\usepackage{graphicx}
\usepackage{xcolor}

\font\japit = cmti10 at 10truept

\title
     [Intergalactic Filaments]
{\vglue-3.0truecm
\centerline{\japit For submission to Monthly Notices}
\vglue 2.5truecm
\noindent
A Model For Intergalactic Filaments and Galaxy Formation During the First Gigayear
\author[A. G. Harford et al.]
	{A. Gayler Harford$^1$ and Andrew J. S. Hamilton$^{1,2}$ \\
	$^1$JILA, University of Colorado, Boulder, CO 80309, USA \\
	$^2$Dept.\ Astrophysical \& Planetary Sciences,
	University of Colorado,
	Box 440,
	Boulder CO 80309, USA}
}

\newcommand{\unit}[1]{\, {\rm #1}}

\newcommand{\boldemp}[1]{#1}

\newcommand{\bold}[1]{#1}

\newcommand{\cmdmodeldiagram}{
    \begin{figure}
    \begin{center}
    \leavevmode
    \includegraphics[scale=0.7]{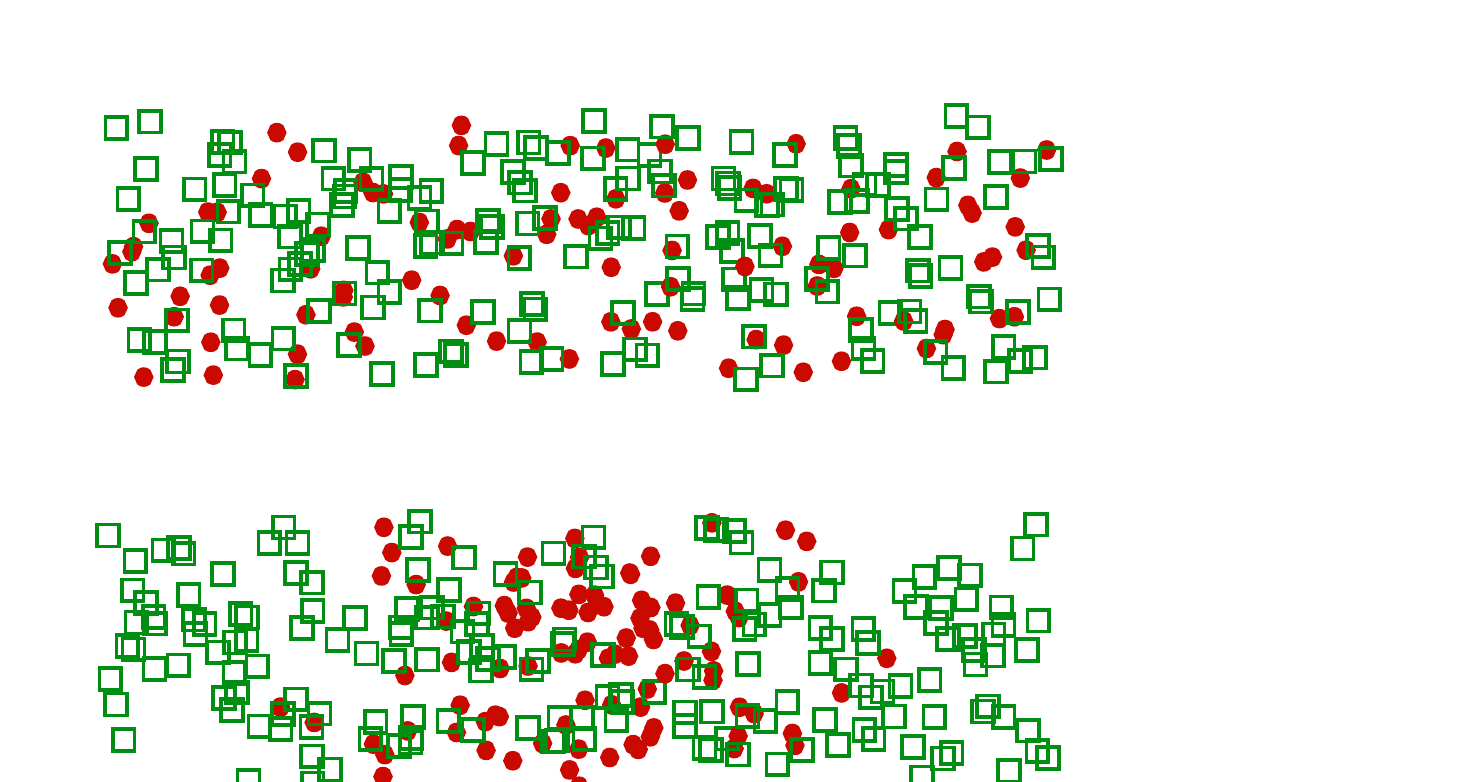}
    \end{center}
    \caption[1]{
    \label{modeldiagram}
    \bold{Schematic diagram of the model.}
    These two images illustrate the model schematically
    in its purest form.  The model is formulated as the collapse
    of a rod to form a quasi-spherical galaxy. Imagine that the
    centre of collapse is in the centre of each image.  
    The upper image shows the rod at an early time
    with dark matter shown as red, filled circles and gas 
    as green squares.  The gas assumes the
    structure of a gravitationally bound, isothermal cylinder.
    The dark matter contributes gravitational effects but not
    hydrodynamic ones.  To facilitate computation we assume
    that the dark matter and gas are uniformly mixed, at least
    initially.
    The lower image shows the rod as it might appear at
    a later time.  Dark matter has begun to collapse
    toward the centre
    while the gas has remained extended because of the
    additional hydrodynamic effects that oppose the
    gravitational coalescence.  In this paper, when the
    model is compared to the simulation, the two halves of
    the rod on either side of the centre of collapse
    are regarded as two separate filaments since both sides
    are not always present for every galaxy.  The structure
    of such a filament is sampled at a position well separated
    from the centre of the forming galaxy.
    An example from the simulation that
    resembles this diagram is shown in Figure~\ref{modelexample}.
    }
    \end{figure}
}

\newcommand{\cmdmodelexample}{
    \begin{figure}
    \begin{center}
    \leavevmode
    \includegraphics[scale=0.4]{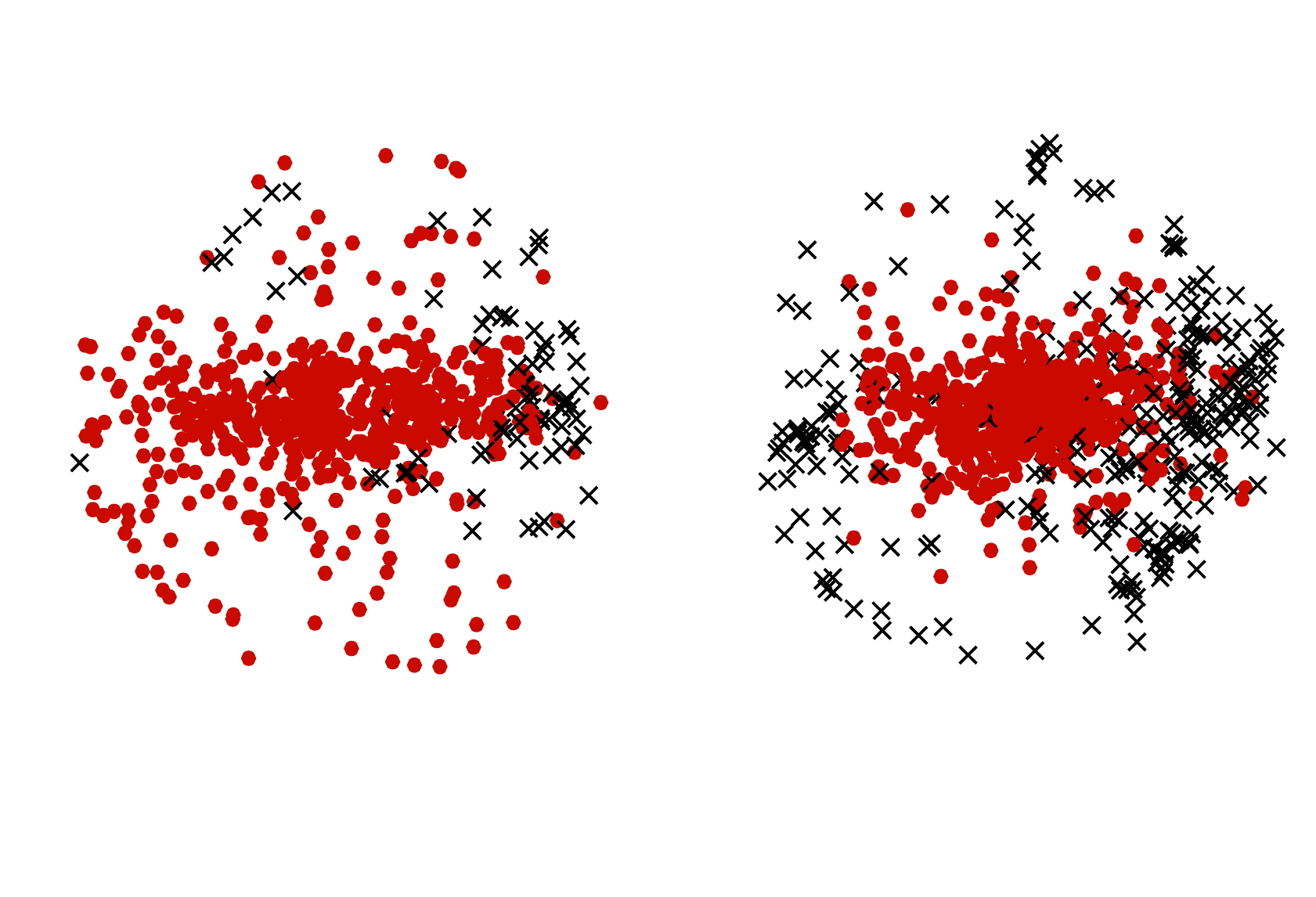}
    \includegraphics[scale=0.4]{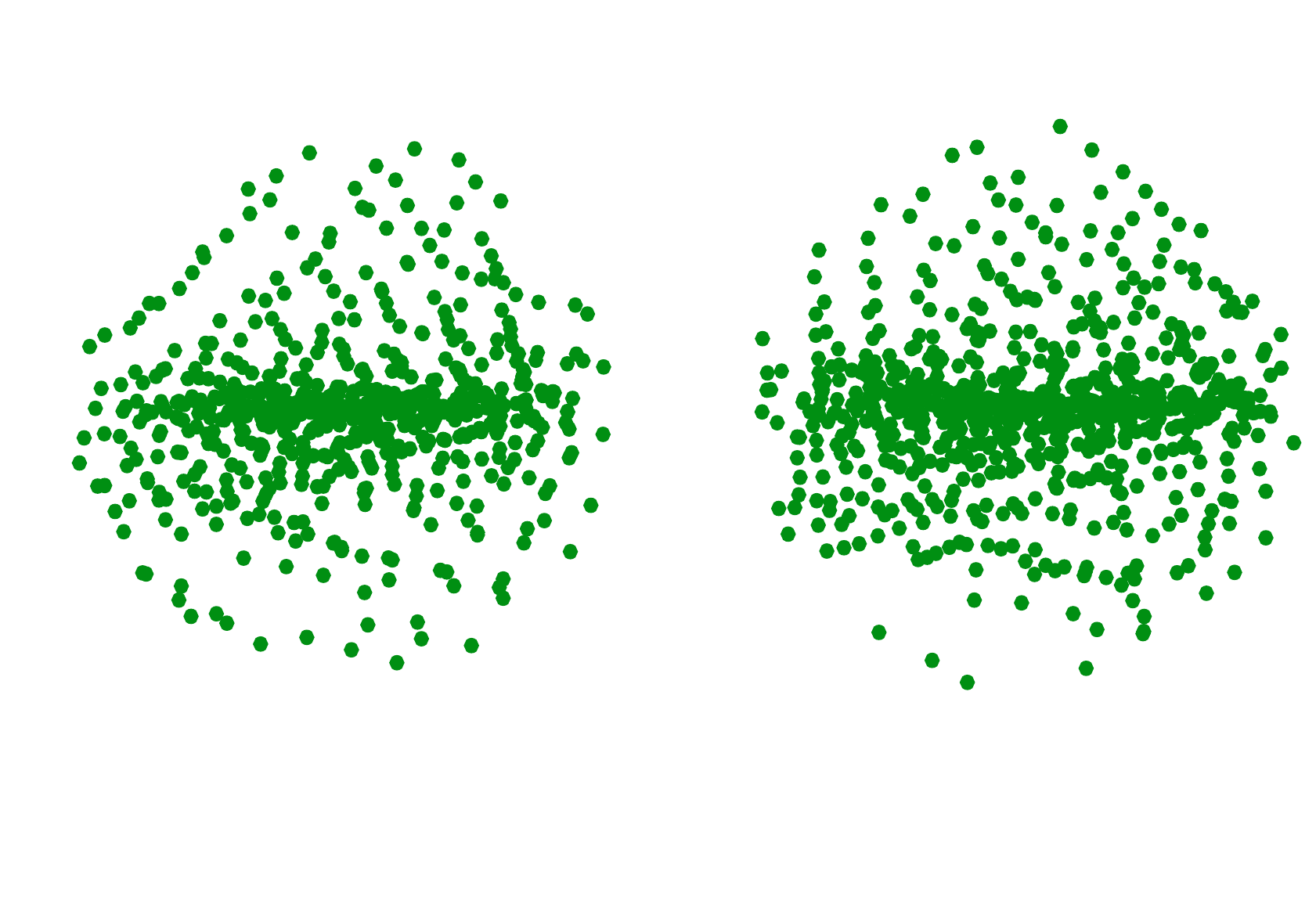}
    \end{center}
    \caption[1]{
    \label{modelexample}
    \bold{Images from the simulation that resemble the schematic
      drawing in Figure~\ref{modeldiagram}.}  These are projections
      of a sphere centred on a single galaxy at an 
      early time (left) 
      and at a later time (right).
      The top two images show just the dark matter.  
      The red, filled circles show particles that
      will actually end up in the galaxy at the end of the
      simulation.  They are seen to coalesce in the centre 
      at a
      later time in the image on the right.
      The black X's are other dark matter
      particles. Some of these may contribute 
      gravitationally to the
      structure of the filament.  The bottom two images show
      just the gas particles as green, filled circles.  
      They  assume a relatively smooth
      filamentary structure throughout.
    }
    \end{figure}
}

\newcommand{\cmdnumpeak}{
    \begin{figure}
    \begin{center}
    \leavevmode
    \includegraphics[scale=0.5]{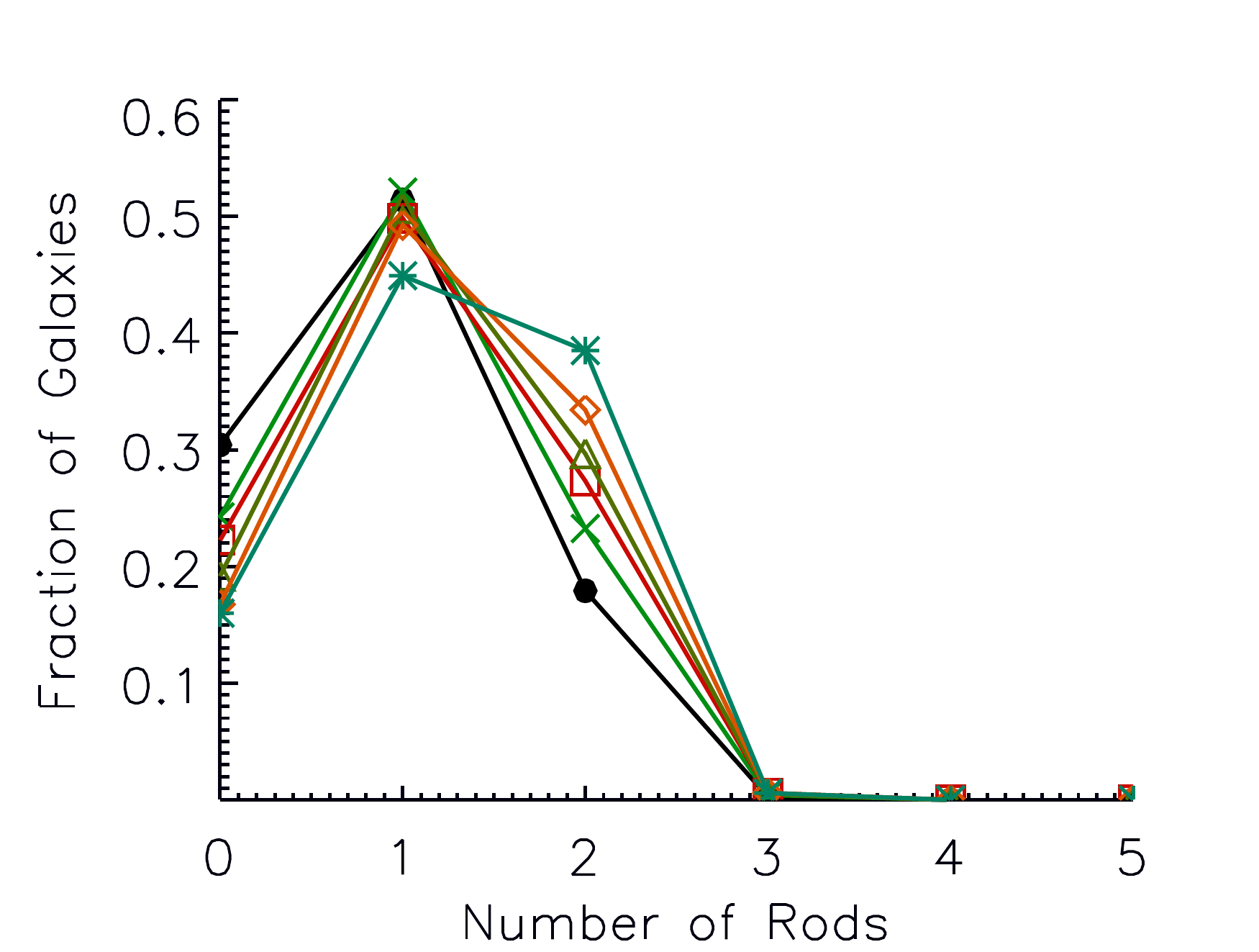}
    \end{center}
    \caption[1]{
    \label{numpeak}
    \bold{Number of rods per galaxy.}
    Data are shown for each of the six redshifts studied:
    $8.90$ (black, filled circles),
    $8.09$ (green X's), $7.33$ (red squares),
    $6.69$ (dull green triangles),
    $5.85$ (orange diamonds), and
    $5.134$ (blue stars).
    }
    \end{figure}
}

\newcommand{\cmdpeakalign}{
    \begin{figure}
    \begin{center}
    \leavevmode
    \includegraphics[scale=0.5]{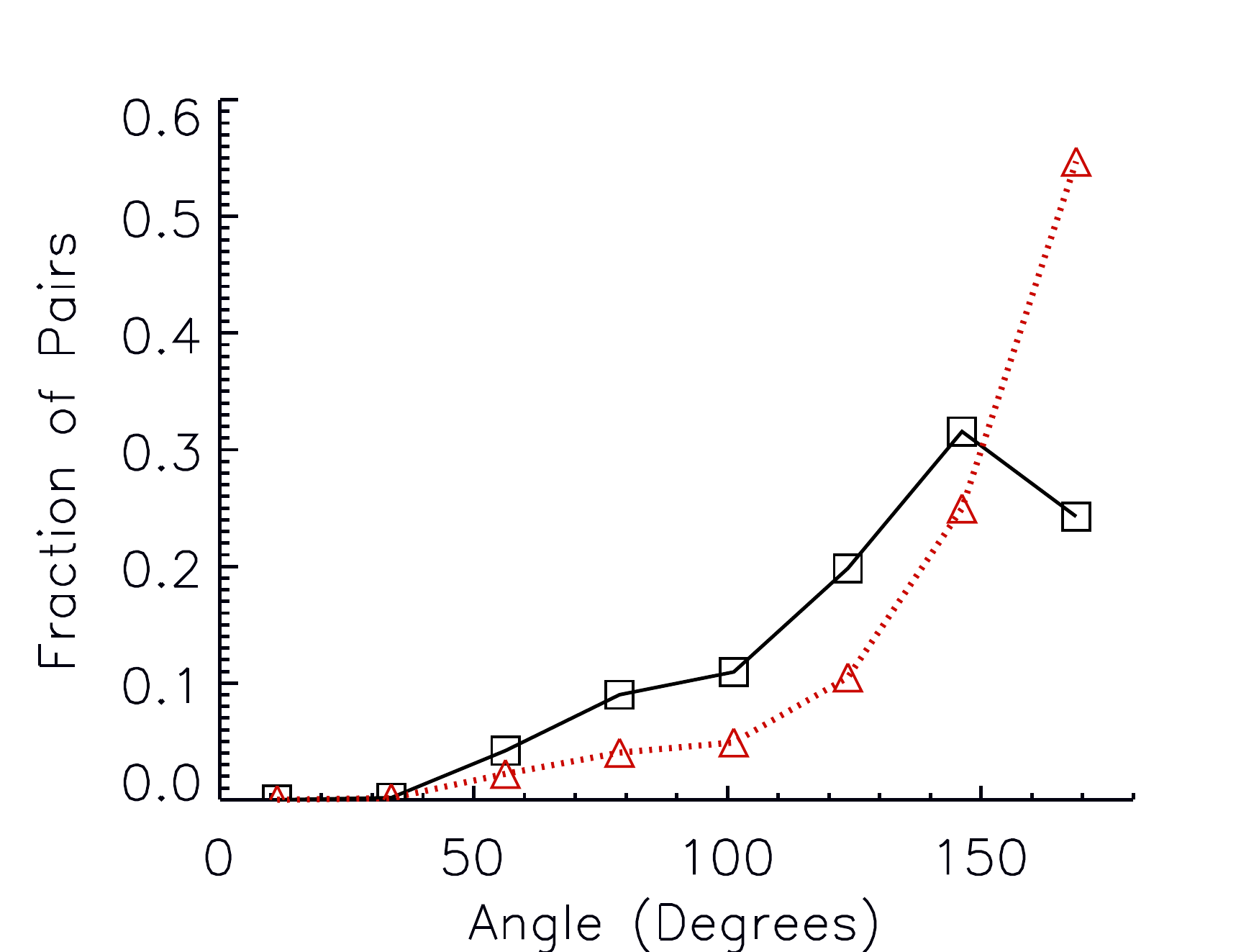}
    \end{center}
    \caption[1]{
    \label{peakalign}
    \bold{Angle between filament segments.}
    Black, solid line with squares shows a histogram of the angle
    between the two segment directions in cases where two are found.
    Red, dotted line with triangles shows the same results when
    the entries have
    been weighted by the reciprocal of the sine of the angle.
    }
    \end{figure}
}

\newcommand{\cmdexcycloid}{
    \begin{figure}
    \begin{center}
    \leavevmode
   \includegraphics[scale=0.5]{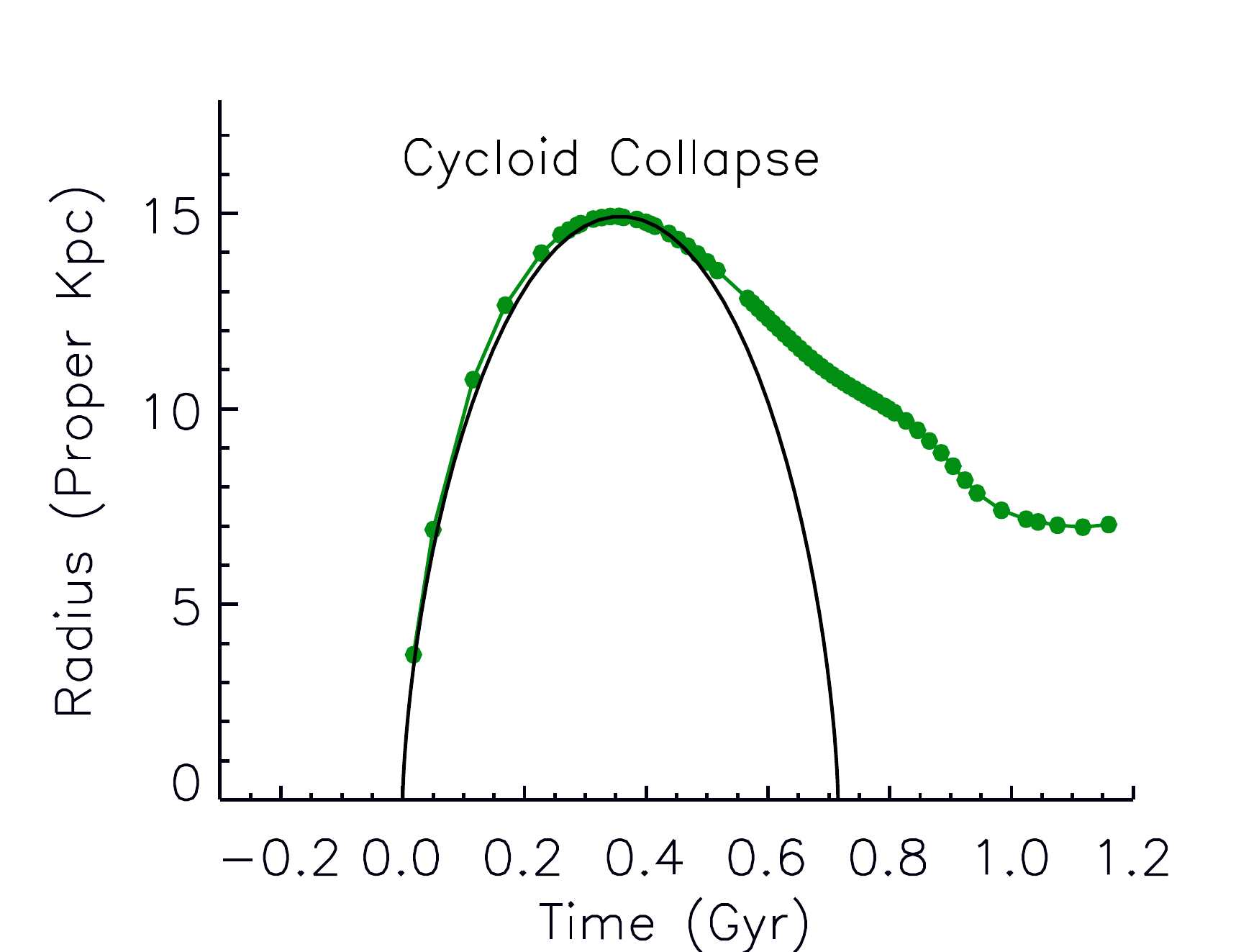}
    \end{center}
    \caption[1]{
    \label{excycloid}
    \bold{Spherical collapse of dark matter in an example galaxy.}
    The average proper radius of galaxy dark matter particles as a
    function of time after the big bang is plotted as a green
    solid line with green filled circles.  The corresponding
    cycloid is shown by the solid black line.
    }
    \end{figure}
}

\newcommand{\cmdoverden}{
    \begin{figure*}
    \begin{center}
    \leavevmode
    \includegraphics[scale=0.36]{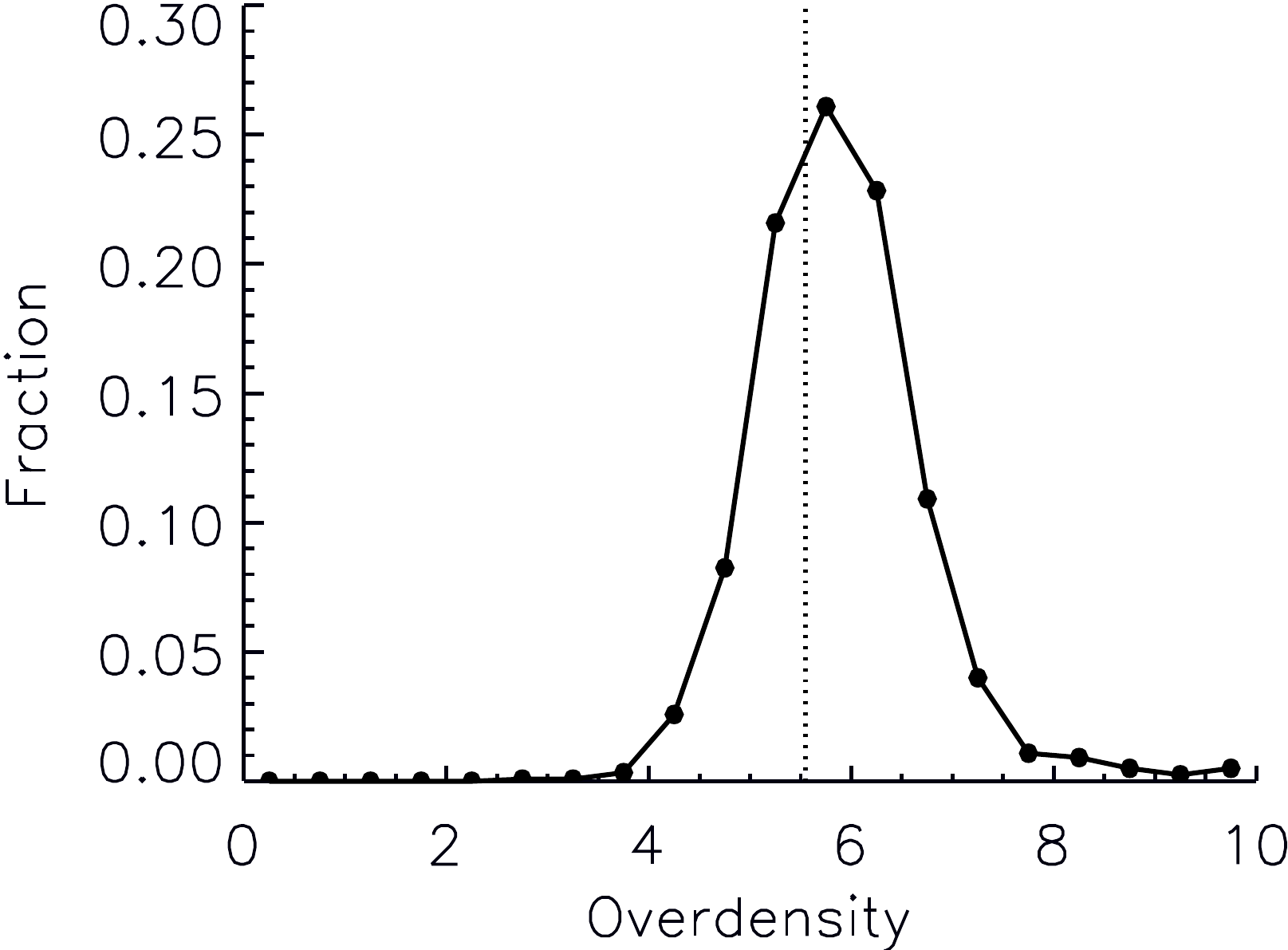}
    \includegraphics[scale=0.36]{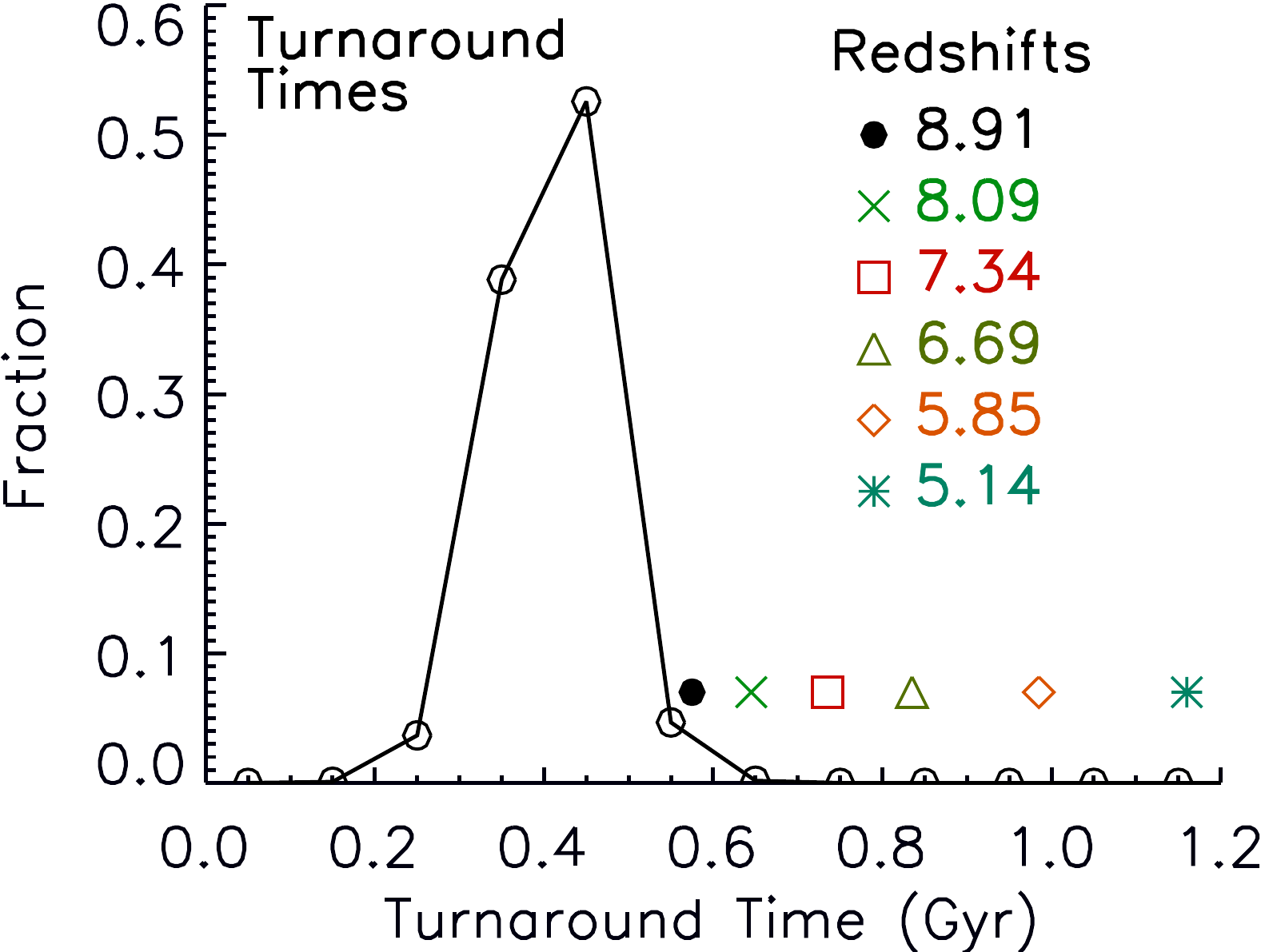}
    \includegraphics[scale=0.36]{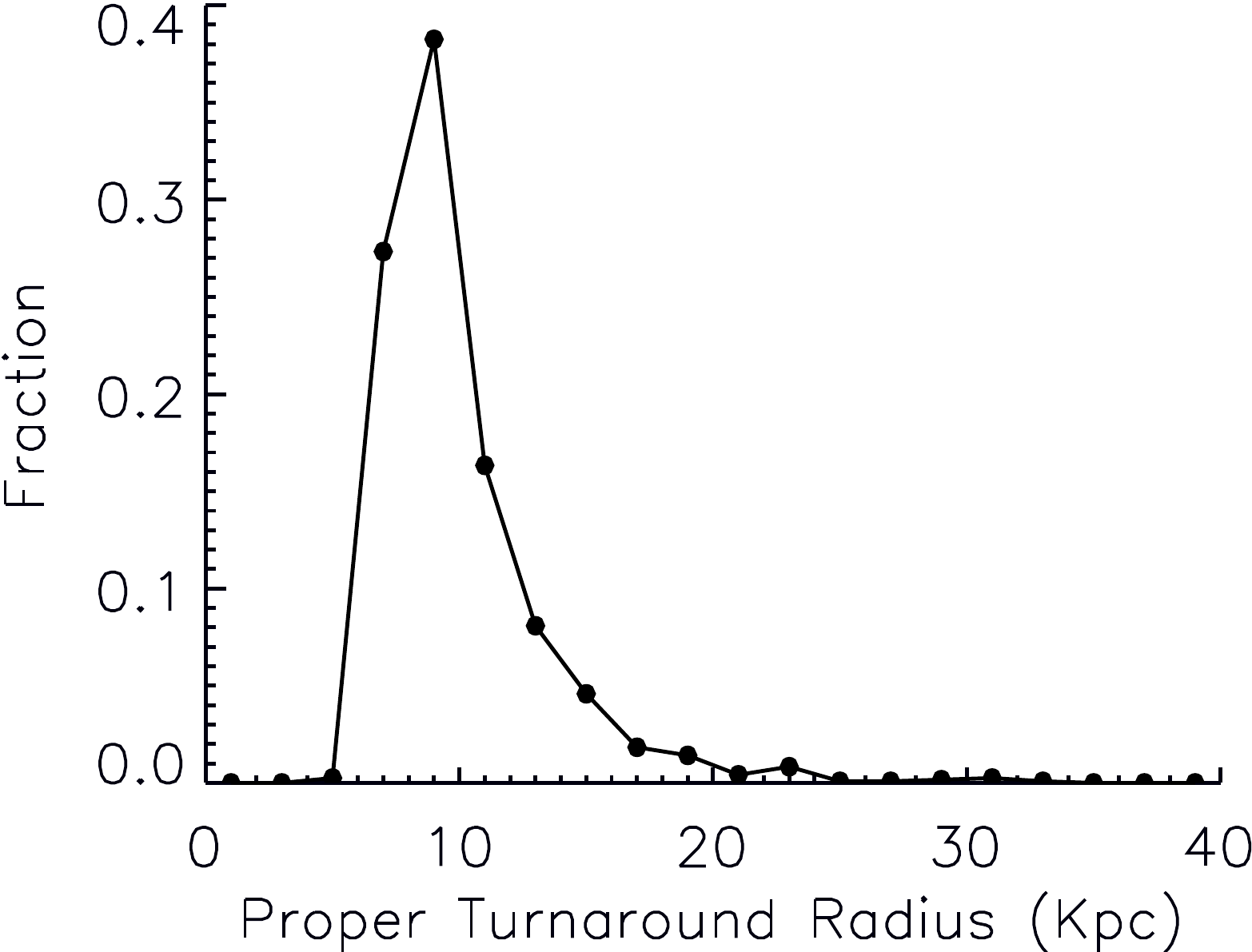}
    \end{center}
    \caption[1]{
    \label{overden}
    Histograms of
    (left) overdensity, (middle) time, and (right) radius
    at turnaround
    for the 1{,}200 largest galaxies in the simulation.
    In the left panel,
    the overdensity is computed for a spherical volume centred
    at the centre of mass of the dark matter particles and
    having a radius equal to the average radius of the dark
    matter particles of the galaxy.
    The theoretically predicted overdensity
    of $5.55$ \citep{peacock99} is shown by the vertical dotted line.
    In the middle graph the times corresponding to the six
    redshifts at which structural analysis of filaments was done
    are shown for comparison to the histogram of turnaround
    times.
    }
    \end{figure*}
}

\newcommand{\cmdsolo}{
    \begin{figure*}
    \begin{center}
    \leavevmode
    \includegraphics[scale=0.33]{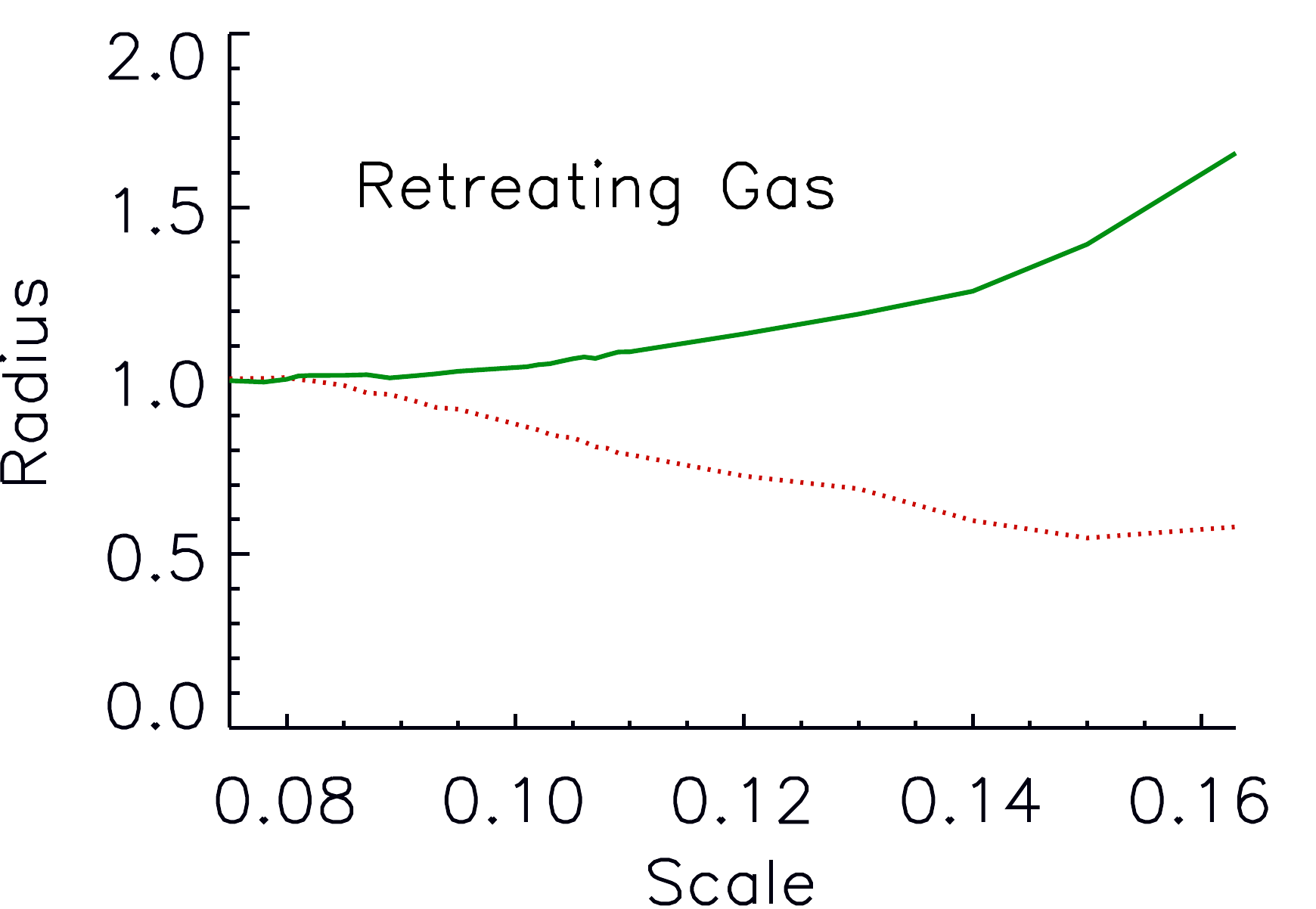}
    \includegraphics[scale=0.33]{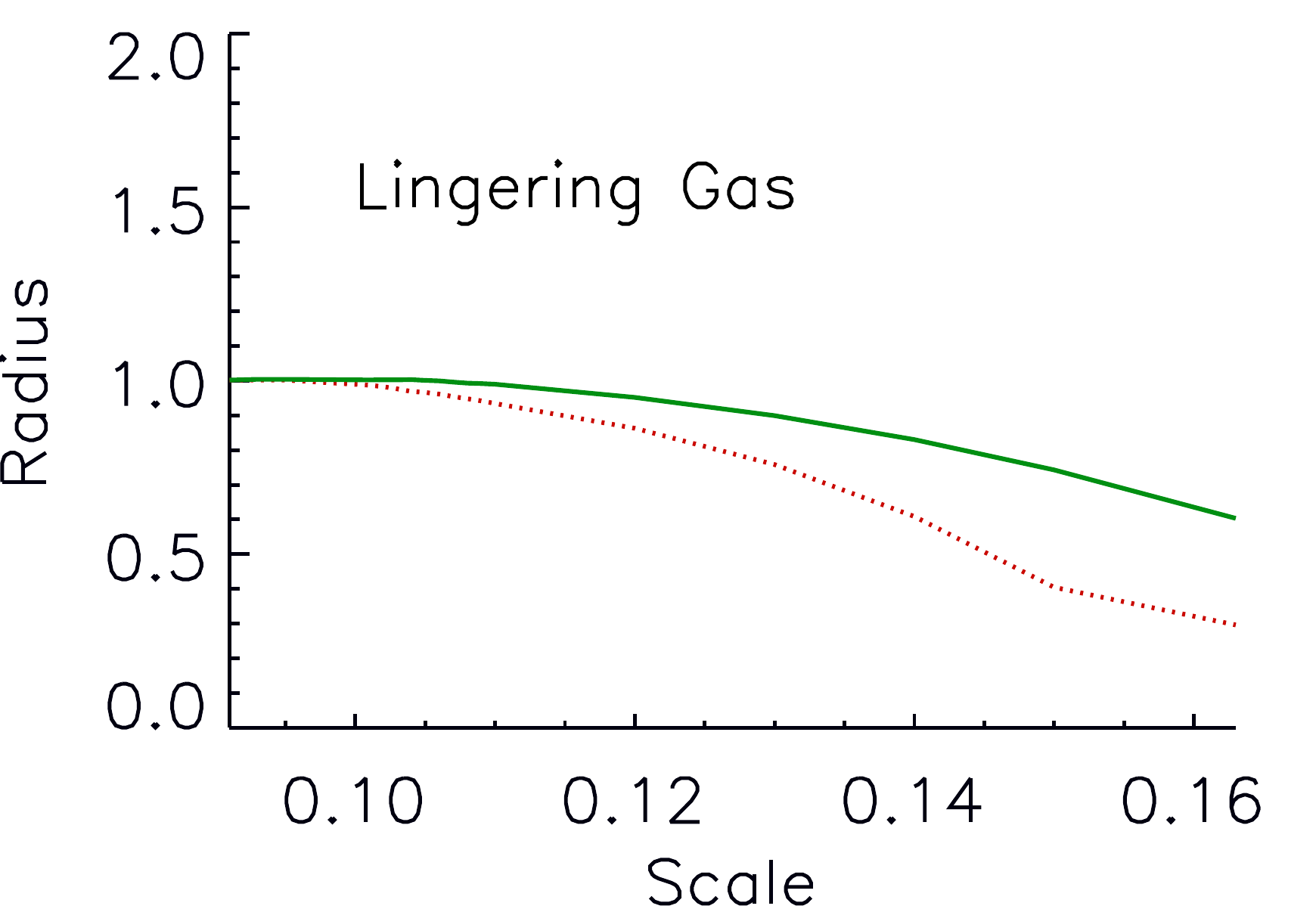}
    \includegraphics[scale=0.33]{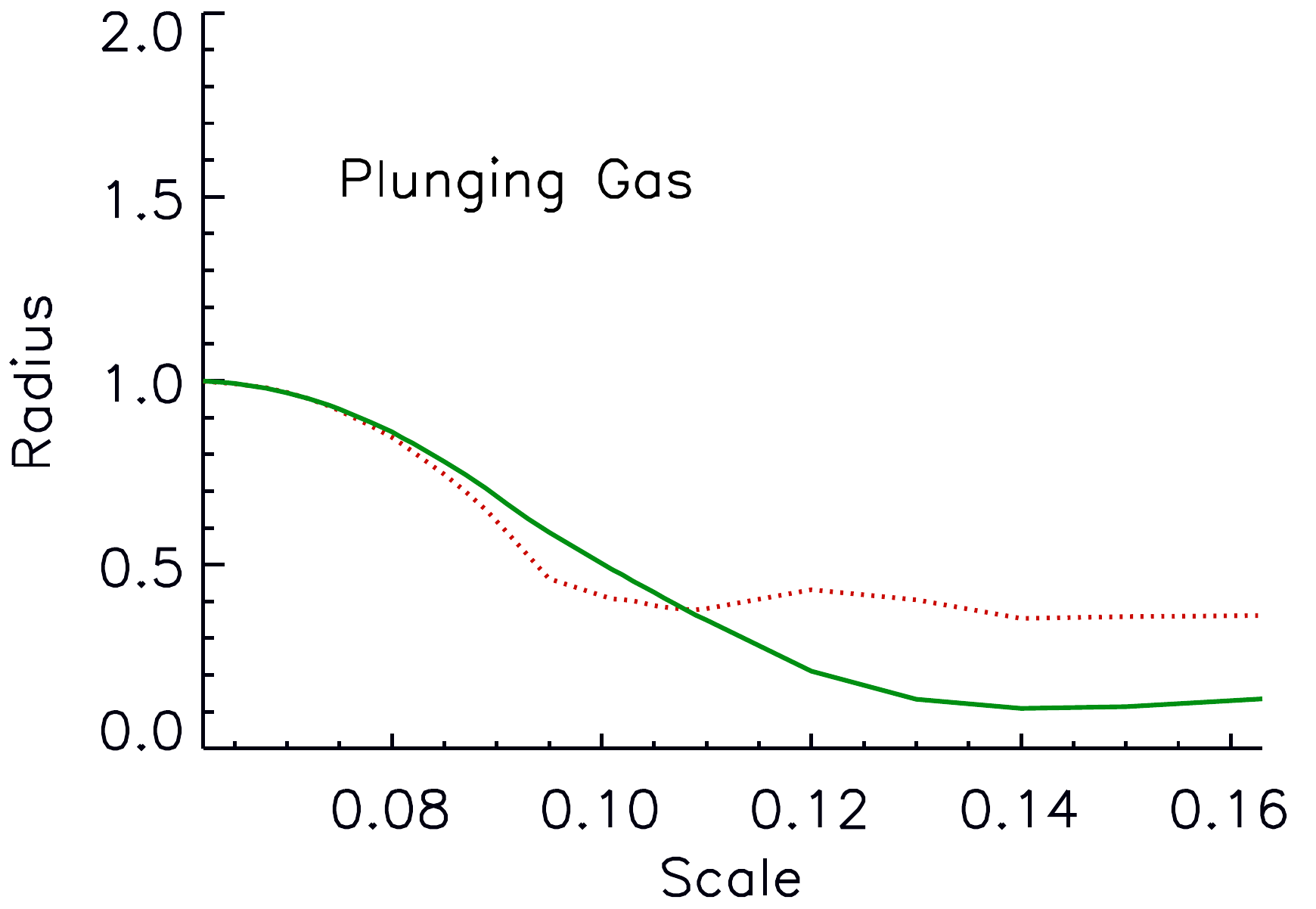}
    \end{center}
    \caption[1]{
    \label{solo}
    \bold{Overall collapse of dark matter and gas in three
    categories of galaxy.}
    Each graph describes a single galaxy chosen to illustrate
    a category.
    Shown is the contrasting motion of dark matter (dotted, red lines)
    and gas (solid green lines) toward the centre of the galaxy.
    At a series of redshifts starting at turnaround the lines show the
    progression of the radius of just that mass of gas or dark matter
    that was within the turnaround radius.  The units of the
    radius are fractions of the turnaround radius.  All radii
    are proper.  The abscissa shows the redshifts as cosmic scale
    factors.  The significance of these categories will be 
    discussed in
    Section~\ref{retard} and ~\ref{sectiontbtf}.
    }
    \end{figure*}
}

\newcommand{\cmdtimescattgfract}{
    \begin{figure}
    \begin{center}
    \leavevmode
    \includegraphics[scale=0.45]{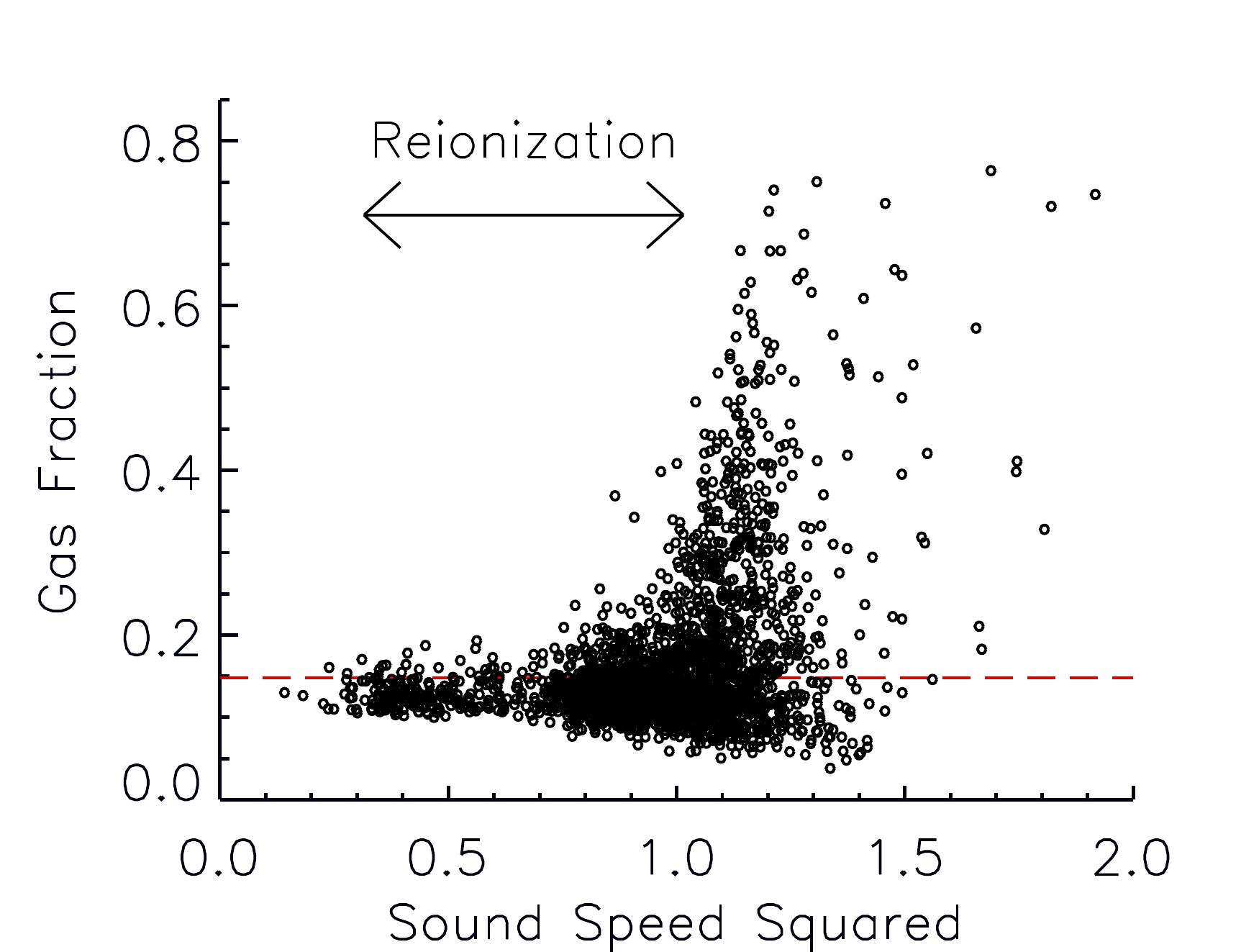}
    \end{center}
    \caption[1]{
    \label{timescattgfract}
    \bold{Gas fraction as function of sound speed.}
    Each black, circle represents an individual
    filament segment. The abscissa is the square of the
    sound speed and the ordinate is the gas fraction.  The arrow
    shows the extent of the period of reionization of the
    filament segments.  The cosmic baryon fraction is
    indicated by the red, dashed line. There is little stellar
    material in the filaments at these times, and its presence
    has been ignored.
    }
    \end{figure}
}

\newcommand{\cmdcollapsegf}{
    \begin{figure}
    \begin{center}
    \leavevmode
    \includegraphics[scale=0.45]{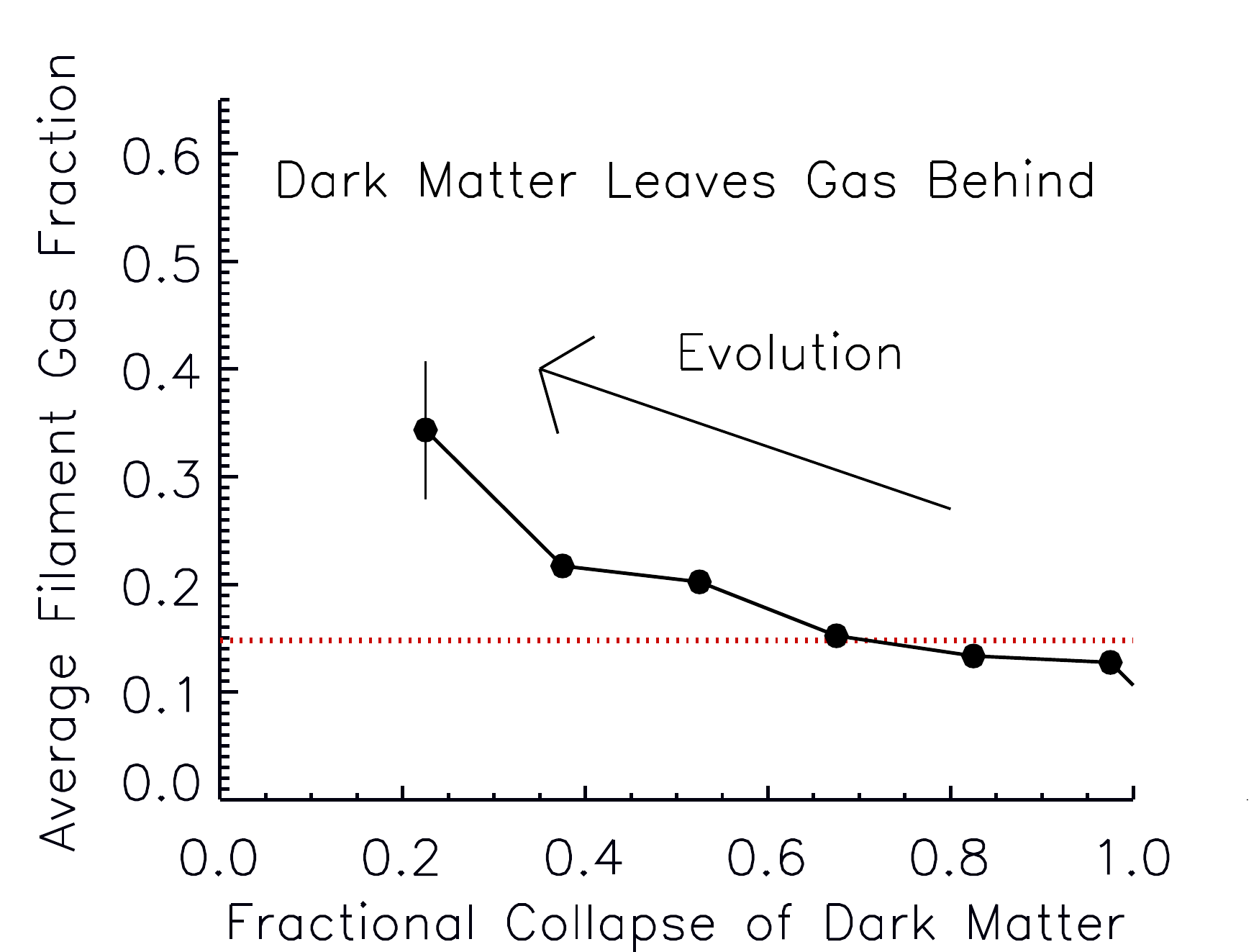}
    \end{center}
    \caption[1]{
    \label{collapsegf}
    \bold{Dark matter leaves gas behind to produce
    baryon enrichment of filaments.}
  The graph shows the average gas fraction
  of filament segments as a function of the collapse fraction
  of the dark matter of the galaxy at the redshift of the segment.
  Collapse fraction is defined and determined as specified in
  Section~\ref{contrast}.  Note that in this figure maximal
  collapse is at the left and minimal collapse at the right.
  The horizontal, dotted, red line indicates the cosmic baryon fraction.
    }
    \end{figure}
}

\newcommand{\cmdbaryrich}{
    \begin{figure}
    \begin{center}    
      \leavevmode
    \includegraphics[scale=0.45]{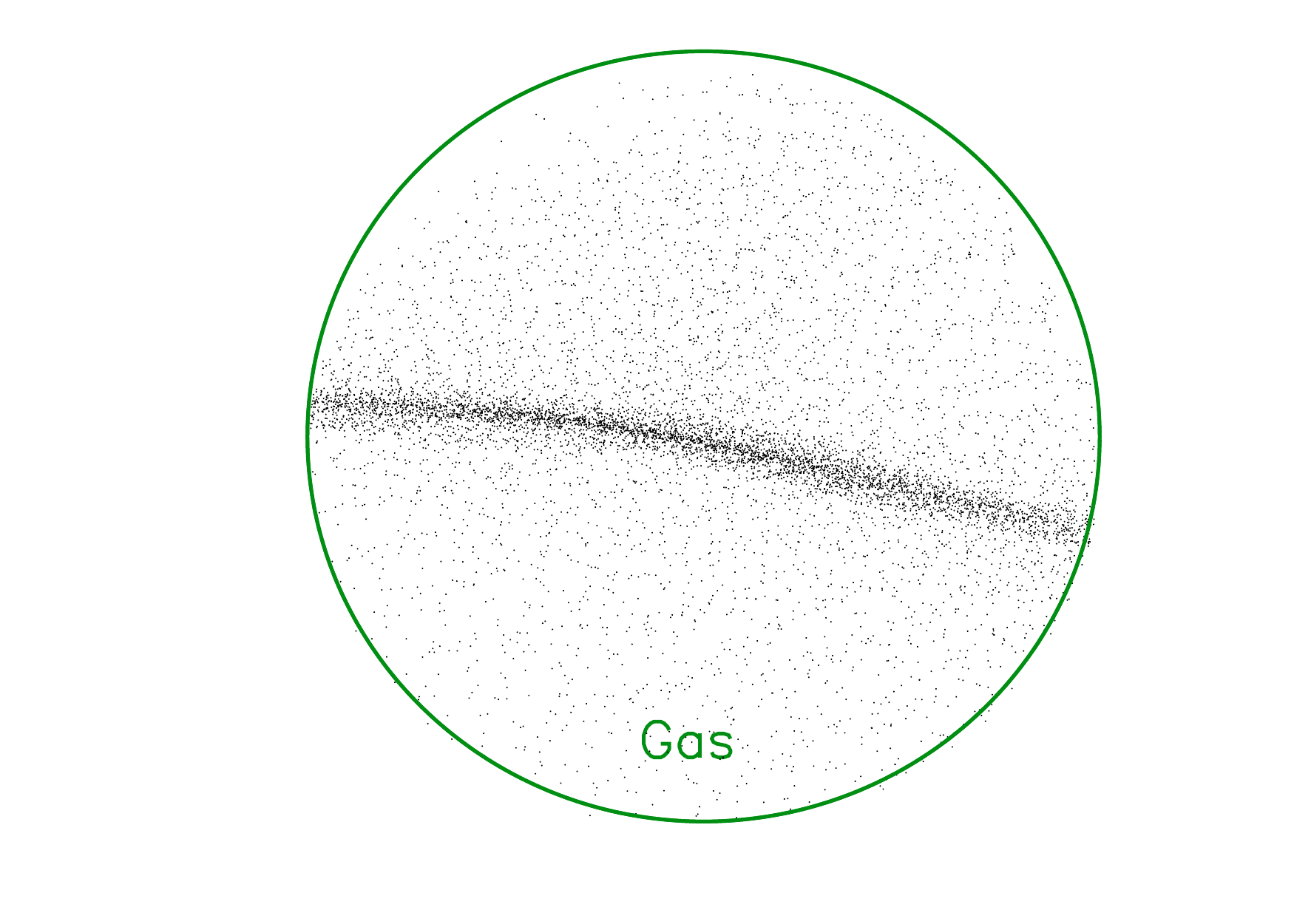}
    \includegraphics[scale=0.45]{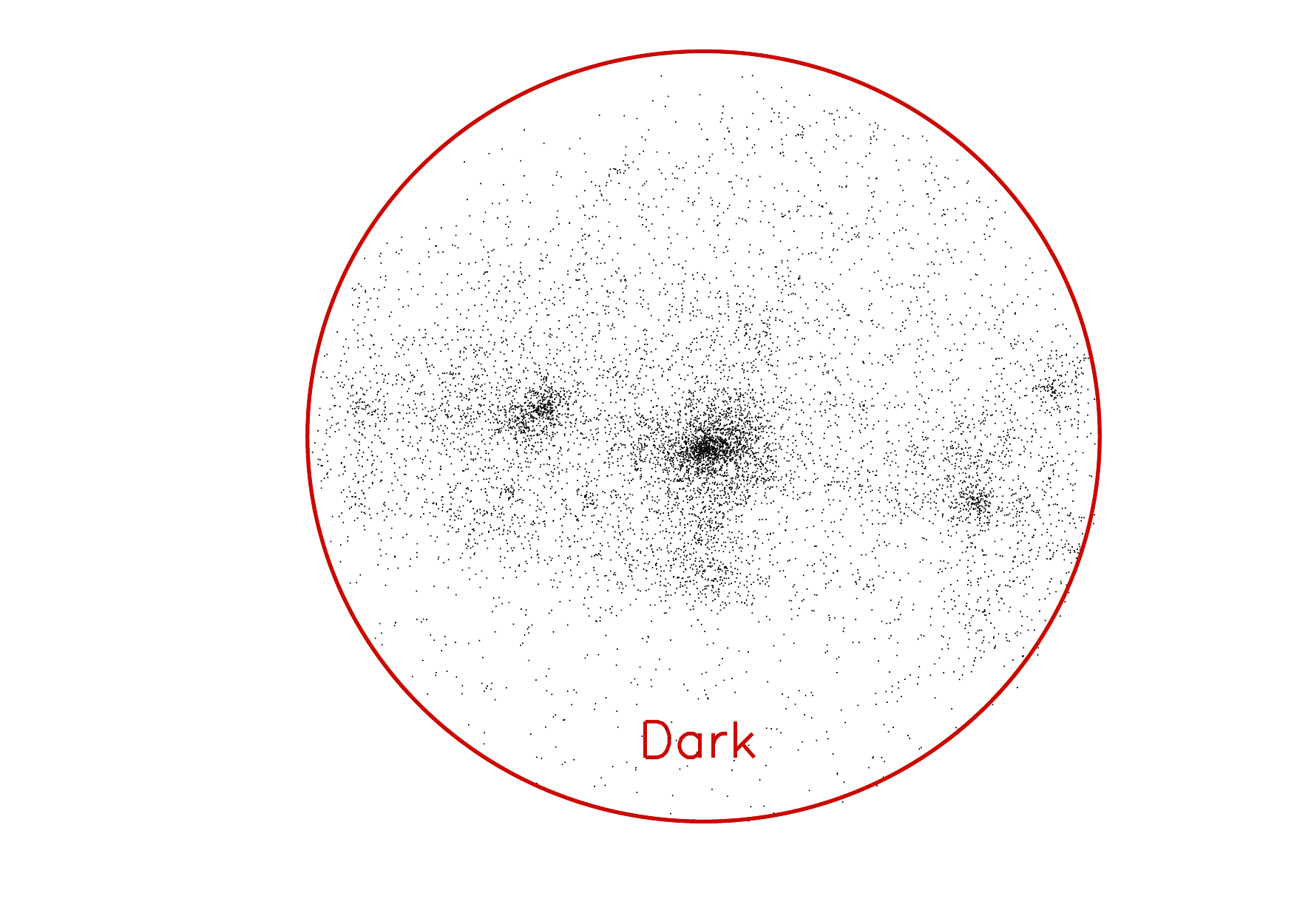}
    \end{center}
    \caption[1]{
    \label{baryrich}
    \bold{Visualization of structures of gas and dark matter.}
    Shown are the gas and dark matter surrounding
    the centre of one of the larger galaxies in the simulation
    at redshift 5.134.
    Gas and dark matter are shown in separate
    images, each of which
    is a projection on to the page of a sphere
    having a comoving radius of $266$ kpc
    centred on the centre of the galaxy.
    The normal to the page
    coincides with the normal to the principal plane of the gas. 
    }
    \end{figure}
}

\newcommand{\cmdreionization}{
    \begin{figure*}
    \begin{center}
    \leavevmode
    \includegraphics[scale=0.30]{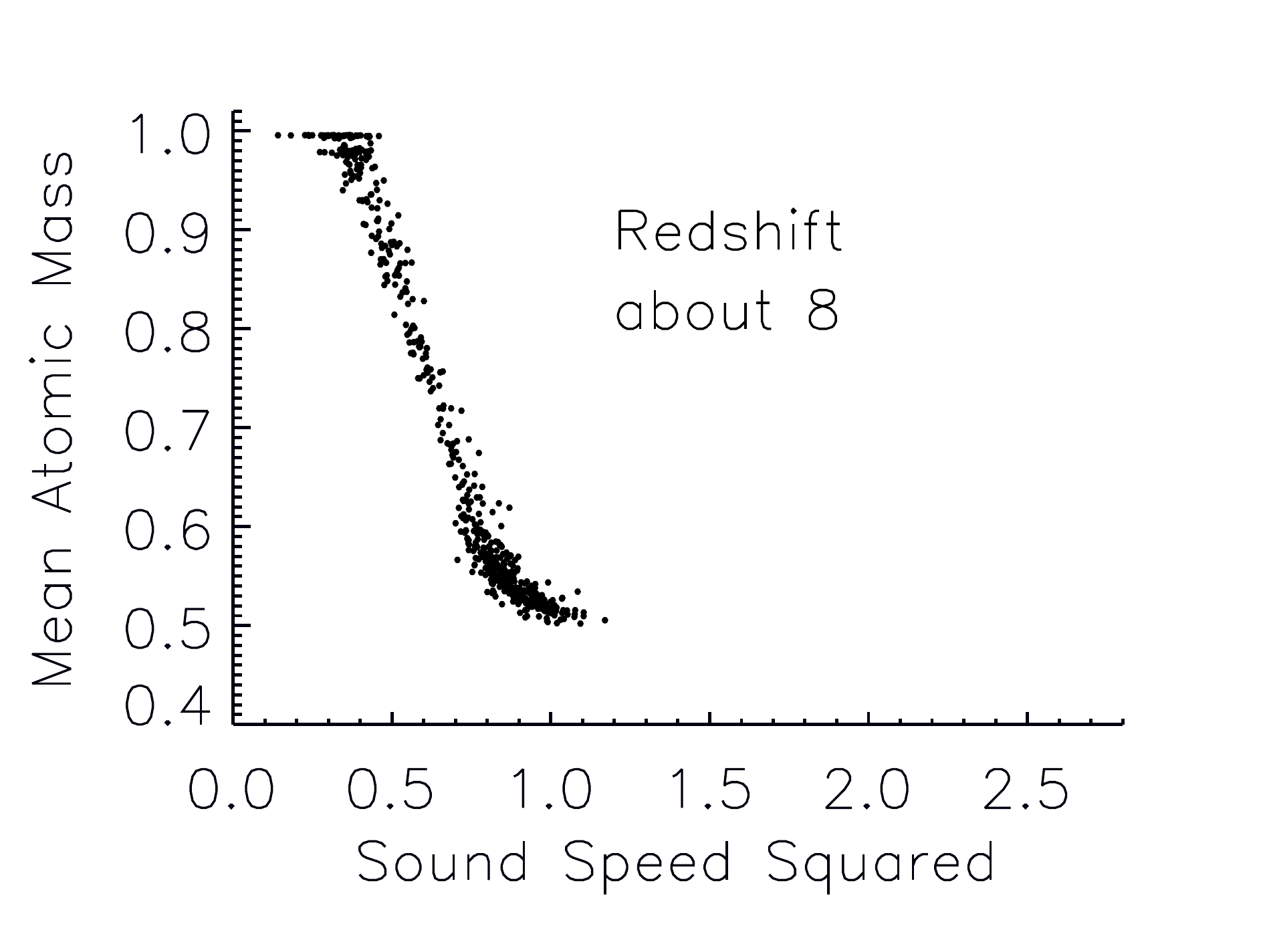}
    \includegraphics[scale=0.30]{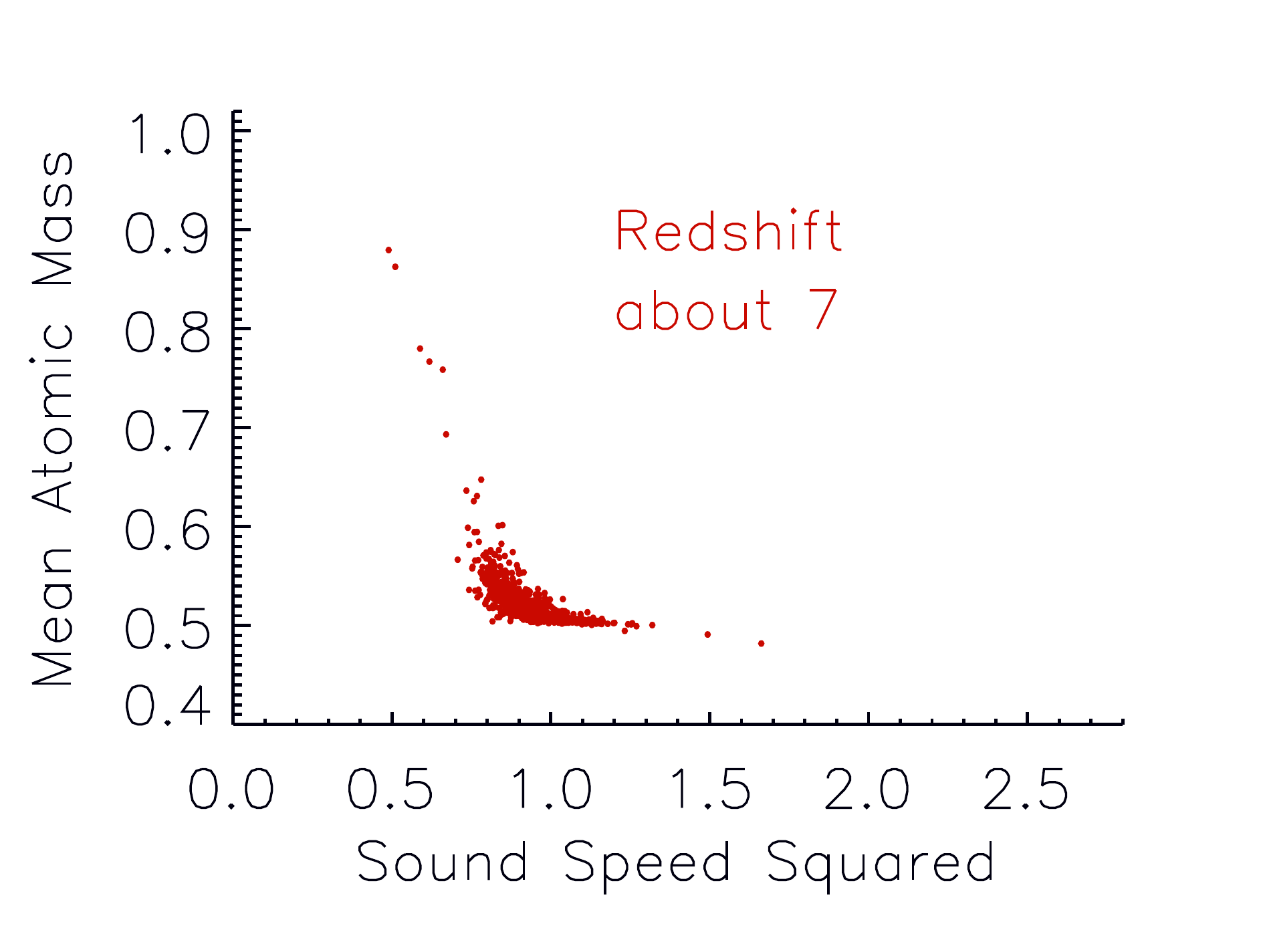}
    \includegraphics[scale=0.30]{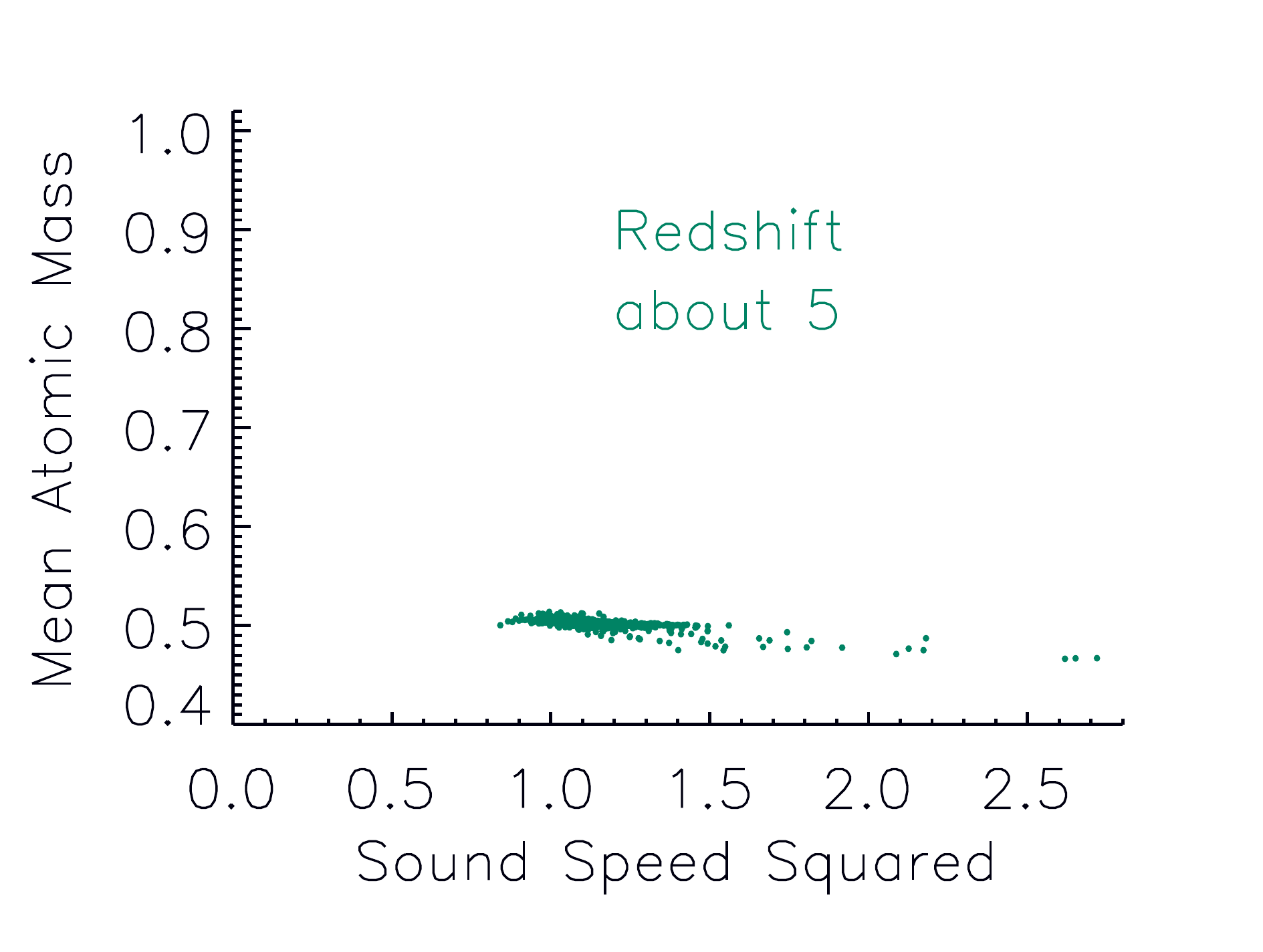}
    \end{center}
    \caption[1]{
    \label{triplet}
    \bold{History of reionization.}
    Scatter plots showing the course of reionization as a
    function of the sound speed squared for three redshift ranges.
    Each dot represents an individual
    filament segment. The abscissa values represent the sound speed
    squared using the arbitrary units of Figure~\ref{figcorr}.
    The ordinate values are mean atomic
    mass ($\mu$ in Equation~\ref{soundspeedequation}).
    The left panel pools the data at redshifts 8.90 and 8.09,
    the middle panel at 7.33 and 6.69, and the right panel at
    5.85 and 5.13.
    }
    \end{figure*}
}

\newcommand{\cmdfigcorr}{
    \begin{figure}
    \begin{center}
    \leavevmode
    \includegraphics[scale=0.45]{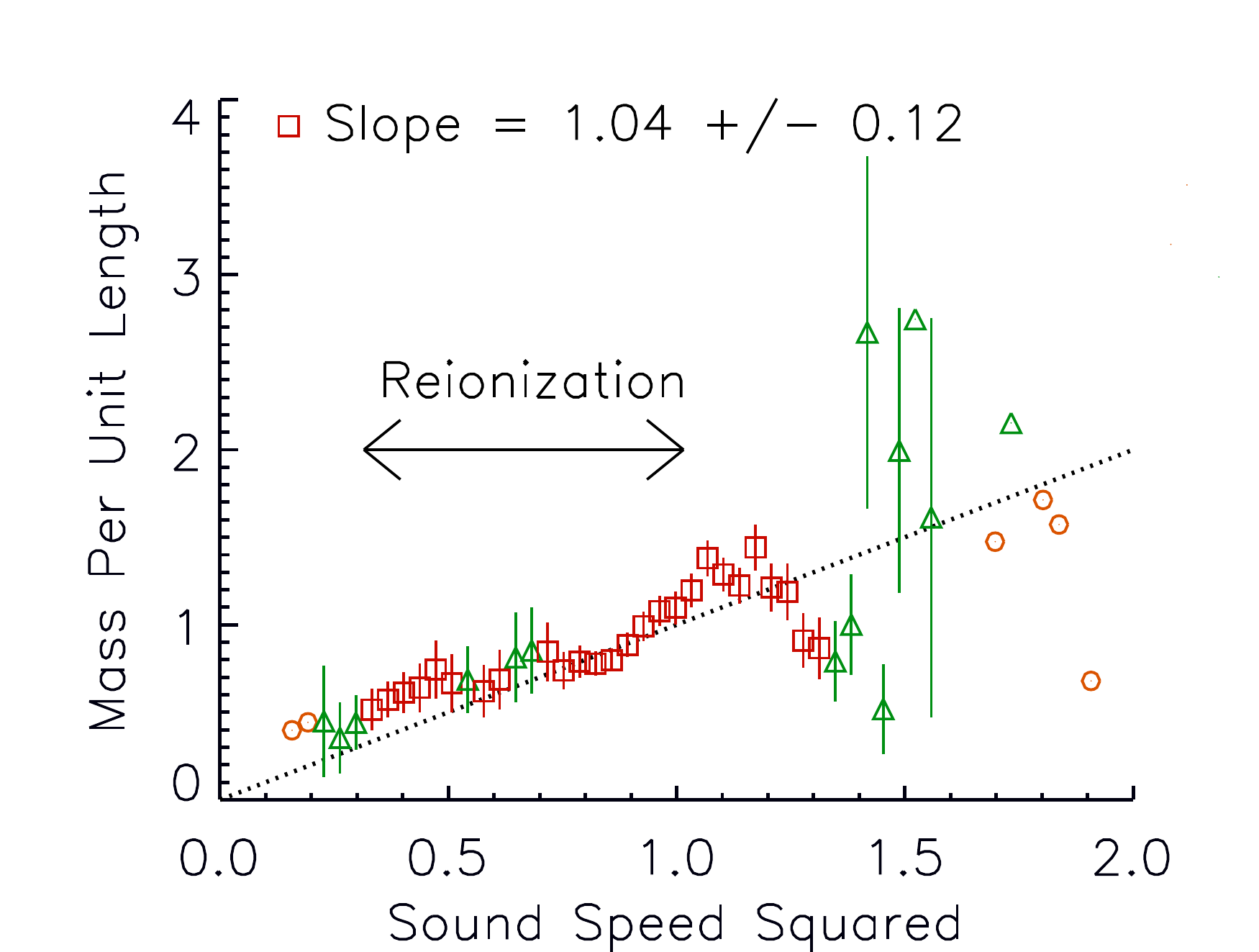}
    \end{center}
    \caption[1]{
    \label{figcorr}
    \bold{Proportionality of total mass per unit length to square of
    sound speed.}
    The symbols show the
    average total (gas plus dark matter) mass
    per unit length
    for the filaments in each bin of sound speed squared.
    The units have been
    chosen to illustrate the expectation of
    proportionality for the model,
    shown as the diagonal, dotted line with a
    slope of one.
    A value of one on the abscissa corresponds to
    a sound speed of $12.9 \unit{km} \unit{s}^{-1}$.
    A value of one on the
    ordinate corresponds to
    $7.71\times10^{7}\unit{M}_\odot \unit{kpc}^{-1}$
    in proper units.
    The bin size is 0.035.
    The vertical line through each symbol represents
    one standard error for the average of
    the filaments in that
    bin.  Standard errors are computed for each bin having
    more than two members.  Each red square
    represents a bin containing at least fifteen filaments.
    The green triangles represent other bins having
    at least
    two filaments, for which a standard deviation can be
    calculated.
    The remaining orange, open
    circles represent bins
    with a single filament.
    The bins with at least fifteen filaments apiece have a linear
    regression line constrained to pass through the origin with a
    slope of $1.04 \pm 0.11$ with a P value of
    $1.7\times10^{-19}$ and RSquared
    of $0.96$.  The probability that the residuals belong to a normal
    distribution is $0.21$.  Regression analysis was
    performed using Mathematica (Wolfram Research).
    }
    \end{figure}
}

\newcommand{\cmdfigcorrb}{
    \begin{figure}
    \begin{center}
    \leavevmode
    \includegraphics[scale=0.45]{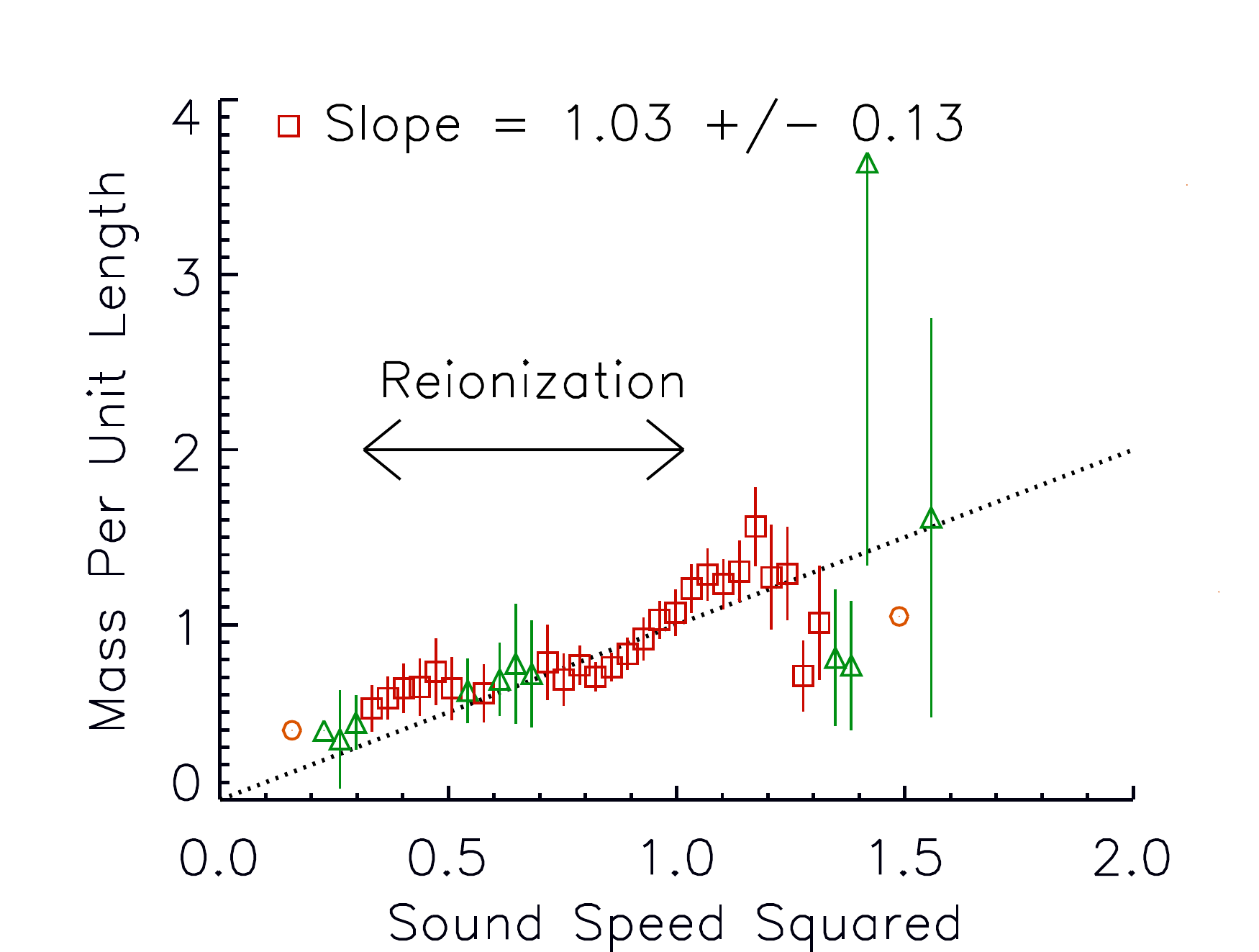}
    \end{center}
    \caption[1]{
    \label{figcorrb}
    \bold{Proportionality of total mass per unit length to square of
    sound speed when controlled for density at turnaround.}
    This figure is the same as Figure~\ref{figcorr} except that the
    filaments have been drawn from a restricted range of proper
    density at turnaround, namely
    $3.0-5.0\times10^{5}\unit{M}_\odot \unit{kpc}^{-3}$.
    The symbols show the
    average total (gas plus dark matter) mass per unit length
    for the filament segments in each bin of sound speed squared.
    The units are the same as described for Figure~\ref{figcorr}.
    The vertical line through each symbol represents
    one standard error for the average of
    the filaments in that
    bin.  Standard errors are computed for each bin having
    more than two members. 
    Each red square
    represents a bin containing at least fifteen segments.
    The green triangles represent other bins having at least
    two segments, for which a standard deviation can be
    calculated.  The remaining orange, open circles represent bins
    with a single segment.
    The bins with at least fifteen segments apiece have a linear
    regression line constrained to pass through the origin with a
    slope of $1.03\pm 0.13$ with a P value of
    $7.1\times10^{-18}$ and RSquared
    of $0.96$.  The probability that the residuals belong to a normal
    distribution is $0.23$.  The regression line is constrained to
    pass through the origin.
    }
    \end{figure}
}

\newcommand{\cmdfigcorrbden}{
    \begin{figure}
    \begin{center}
    \leavevmode
    \includegraphics[scale=0.45]{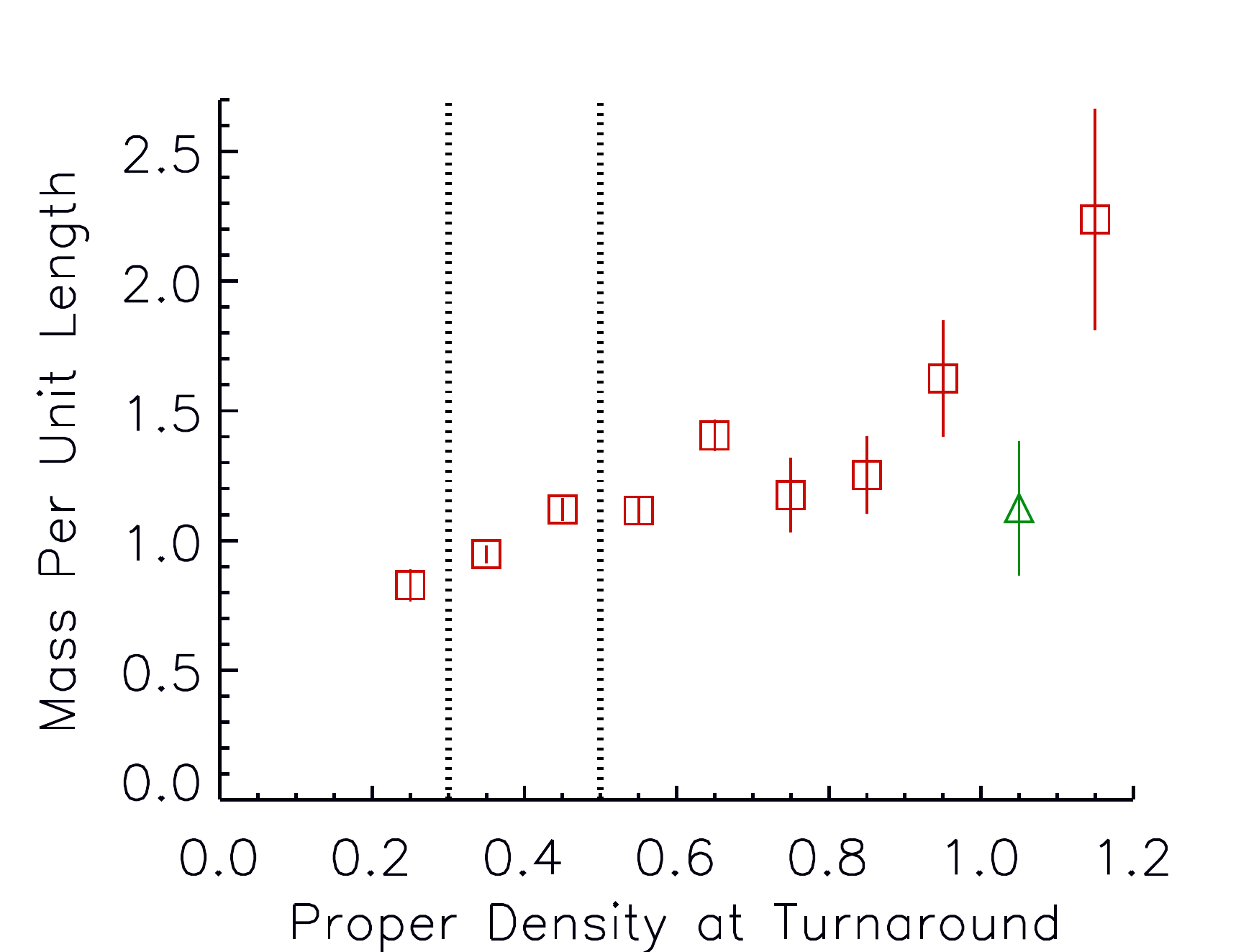}
    \end{center}
    \caption[1]{
    \label{figcorrbden}
    \bold{Total mass per unit length as a function of
     proper density at turnaround.}
    Proper density at turnaround on the abscissa is in units of
    $10^{6}\unit{M}_\odot \unit{kpc}^{-3}$.
    Total mass per unit proper length
    of the filament segment
    on the ordinate is in the same units as for
    Figure~\ref{figcorr} and the symbols are as described
    for that figure.
    Single standard errors are shown by vertical lines.
    The vertical, dotted
    lines indicate the range of turnaround densities used
    to create Figure~\ref{figcorrb}
    }
    \end{figure}
}

\newcommand{\cmdgeom}{
    \begin{figure}
    \begin{center}
     \leavevmode
    \includegraphics[scale=0.45]{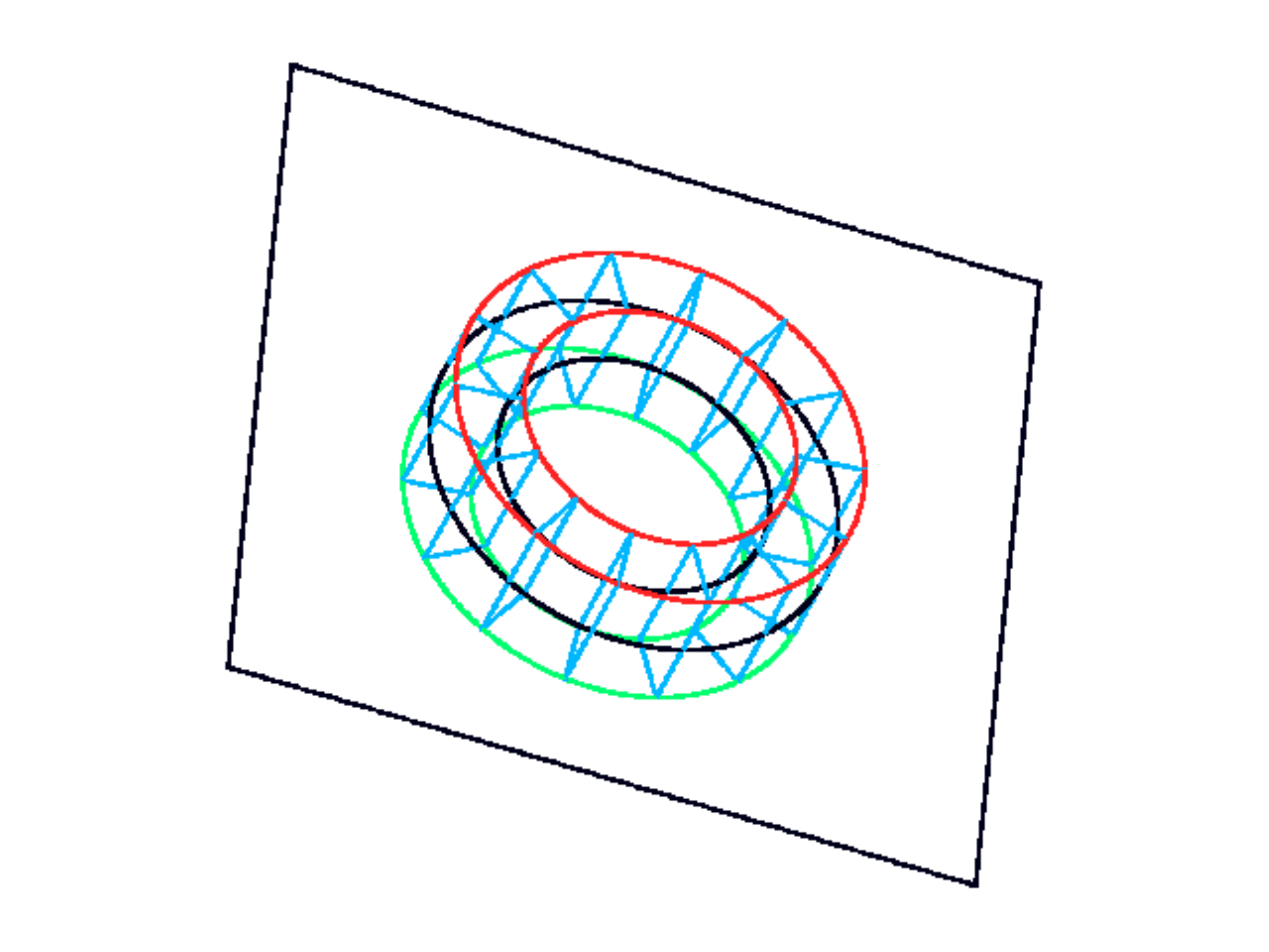}
    \end{center}
    \caption[1]{
    \label{geom}
    \bold{Coordinate system for identification of intergalactic
      filaments.}
    Outlined in black is the XY plane
    with the origin of the
    coordinate system at its centre.  One pair of
    annuli is shown with members above (red) and below (green)
    the central plane,
    The volume between the two annuli is divided into
    equal, azimuthal bins.  The intersection of the
    volume with the plane is shown in black.
    Planes marking the boundaries of the bins are outlined in
    light blue.
    }
    \end{figure}
}

\newcommand{\cmdfilselect}{
    \begin{figure}
    \begin{center}
      \leavevmode
    \includegraphics[scale=0.45]{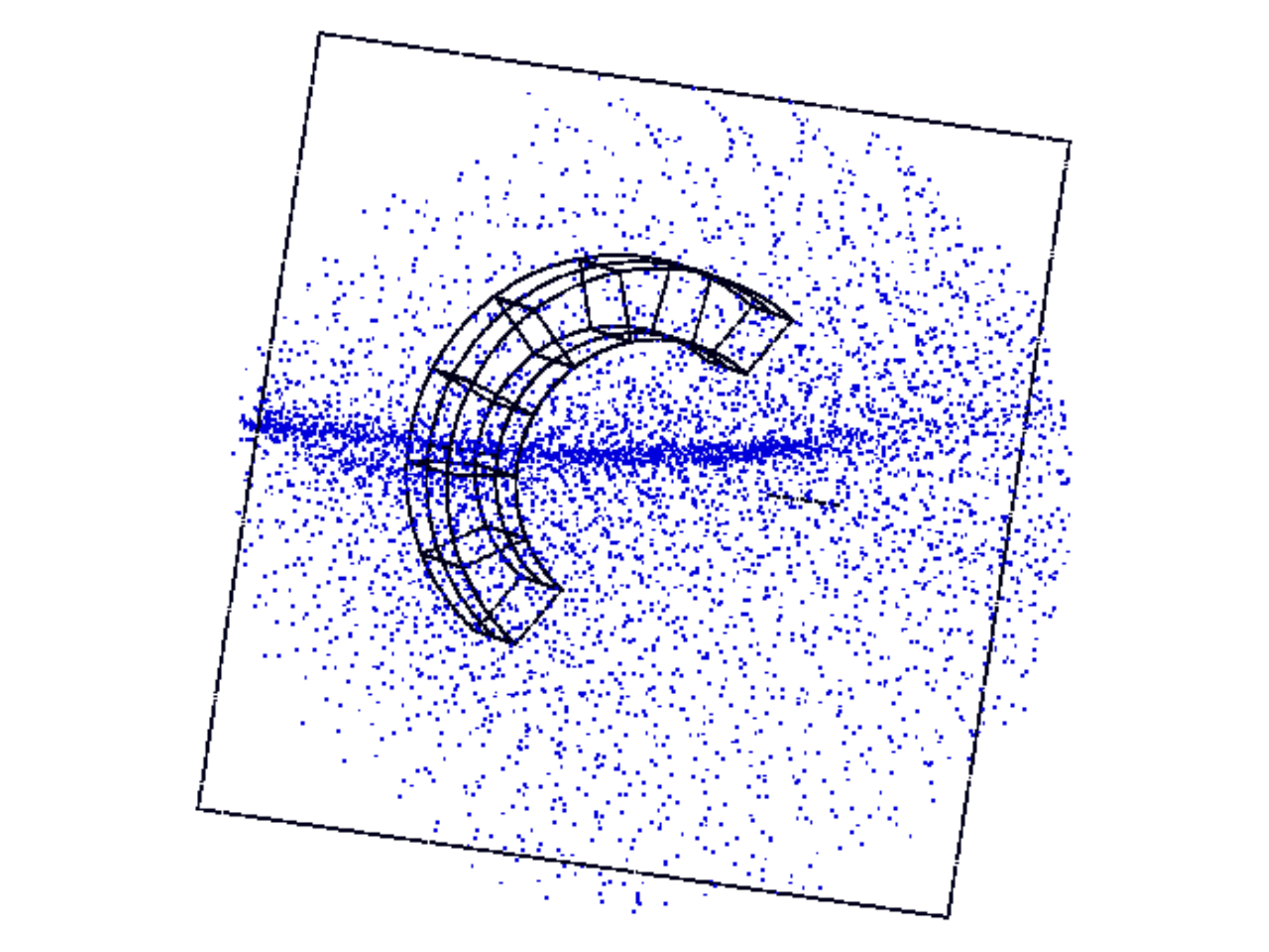}
    \end{center}
    \caption[1]{
    \label{filselect}
    \bold{Identification of a filament extending from the
     centre of a galaxy.}  The centre of the collapsing dark matter
   is at the centre of the coordinate system shown in
   Figure~\ref{geom}, and the plane shown in that figure is
   oriented to the plane of the gas.
   The gas particles in a region
   surrounding the centre of a collapsing galaxy are shown
   in blue.
   For clarity, only some of the bins are shown.
    }
    \end{figure}
}

\newcommand{\cmdbaryturncat}{
    \begin{figure}
    \begin{center}
      \leavevmode
    \includegraphics[scale=0.45]{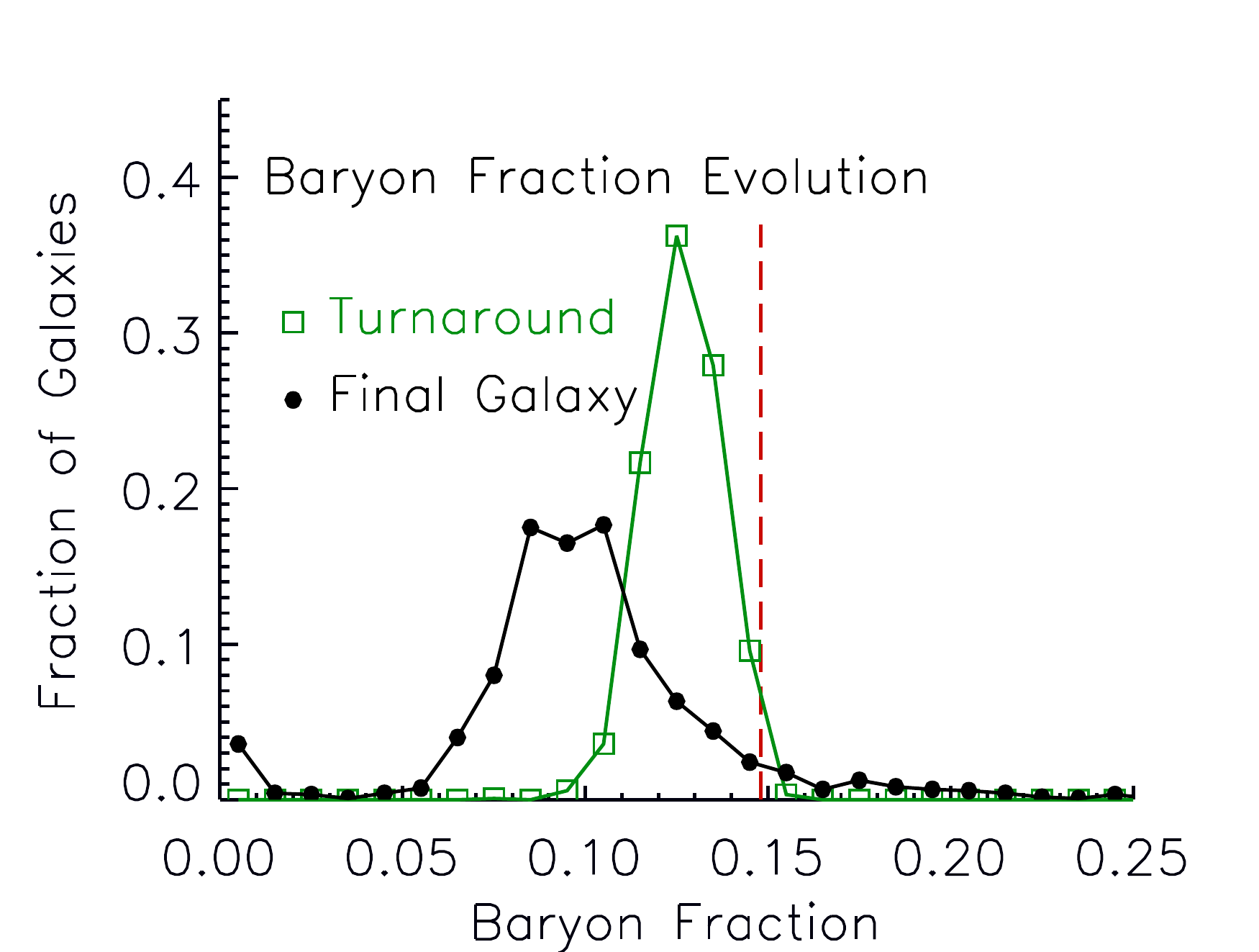}
    \end{center}
    \caption[1]{
      \label{baryturncat}
      \bold{Baryon fraction at turnaround and in final galaxy.}
      Histograms compare the baryon fraction at turnaround
      with that of the final galaxy for all of the galaxies
      in the present study.  The green line with squares represents
      the baryon fraction of the
      matter within the turnaround radius.   The black line with
      filled circles represents the  baryon fraction of the
      final galaxies at the end of the simulation as
      determined by the galaxy finding algorithm.  The
      vertical, red, dashed line indicates the cosmic
      baryon fraction.
    }
    \end{figure}
}

\newcommand{\cmdbfractpredictcube}{
    \begin{figure}
    \begin{center}
      \leavevmode
    \includegraphics[scale=0.45]{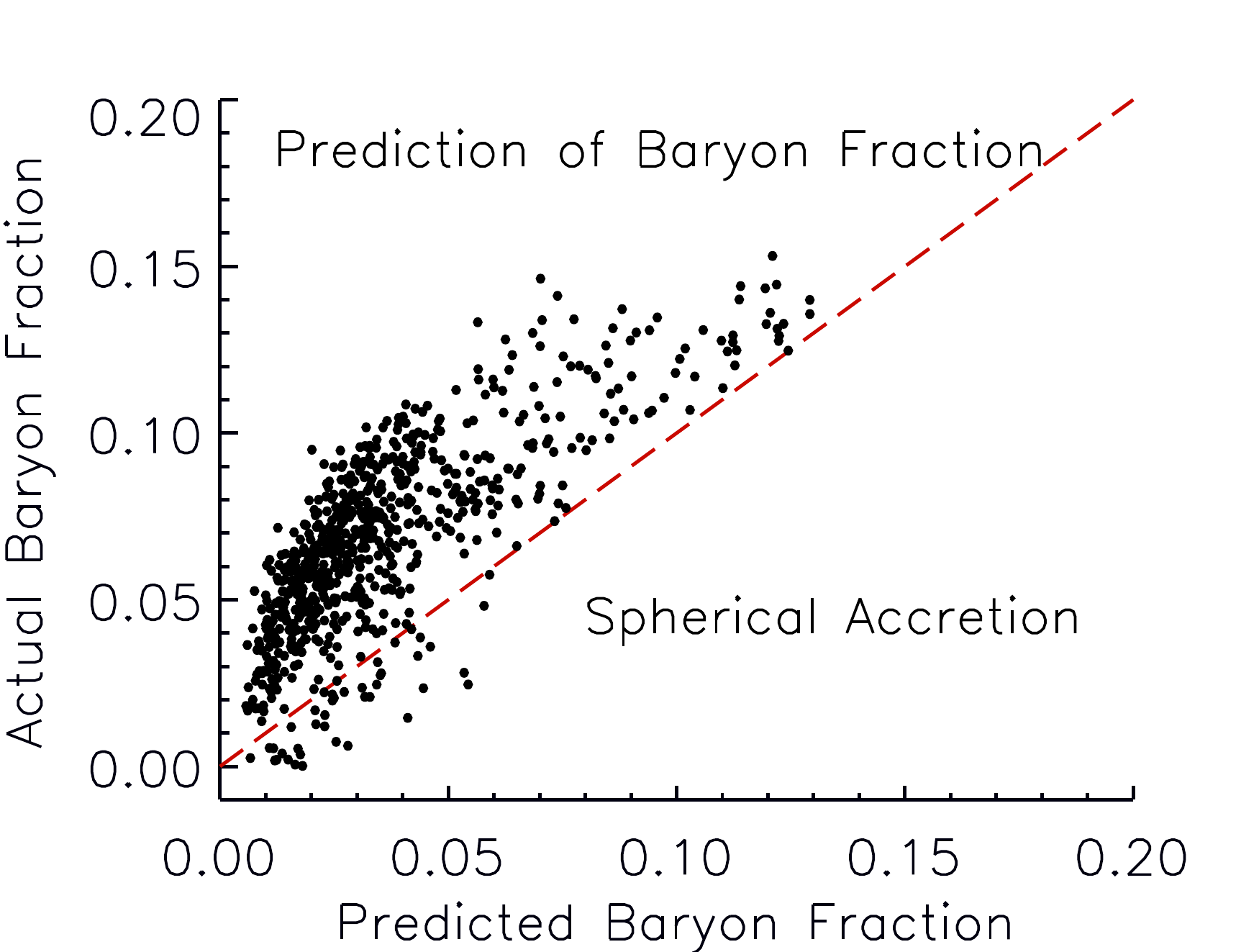}
    \end{center}
    \caption[1]{
      \label{bfractpredictcube}
      \bold{Predicting baryon fraction assuming spherical
        accretion.
      }
      As described in Section~\ref{contrast}, the baryon fraction
      of the final collapse sphere for the dark matter
      is predicted from the baryon fraction
      at turnaround and the final collapse fractions of the dark
      matter and gas.  The computation assumes spherical
      accretion of the gas onto the galaxy.  Each point
      represents a single galaxy.  The abscissa shows the
      predicted value and the ordinate the actual one.
      The dashed, red line represents agreement of the
      prediction with the actual.
      The $5.4$\% of
      galaxies for which the dark matter does not collapse
      faster than the gas are excluded, as well as the galaxies
      where the gas, unlike the dark matter, does not turn
      around at all.
    }
    \end{figure}
}

\newcommand{\cmdbfractpredict}{
    \begin{figure}
    \begin{center}
      \leavevmode
    \includegraphics[scale=0.45]{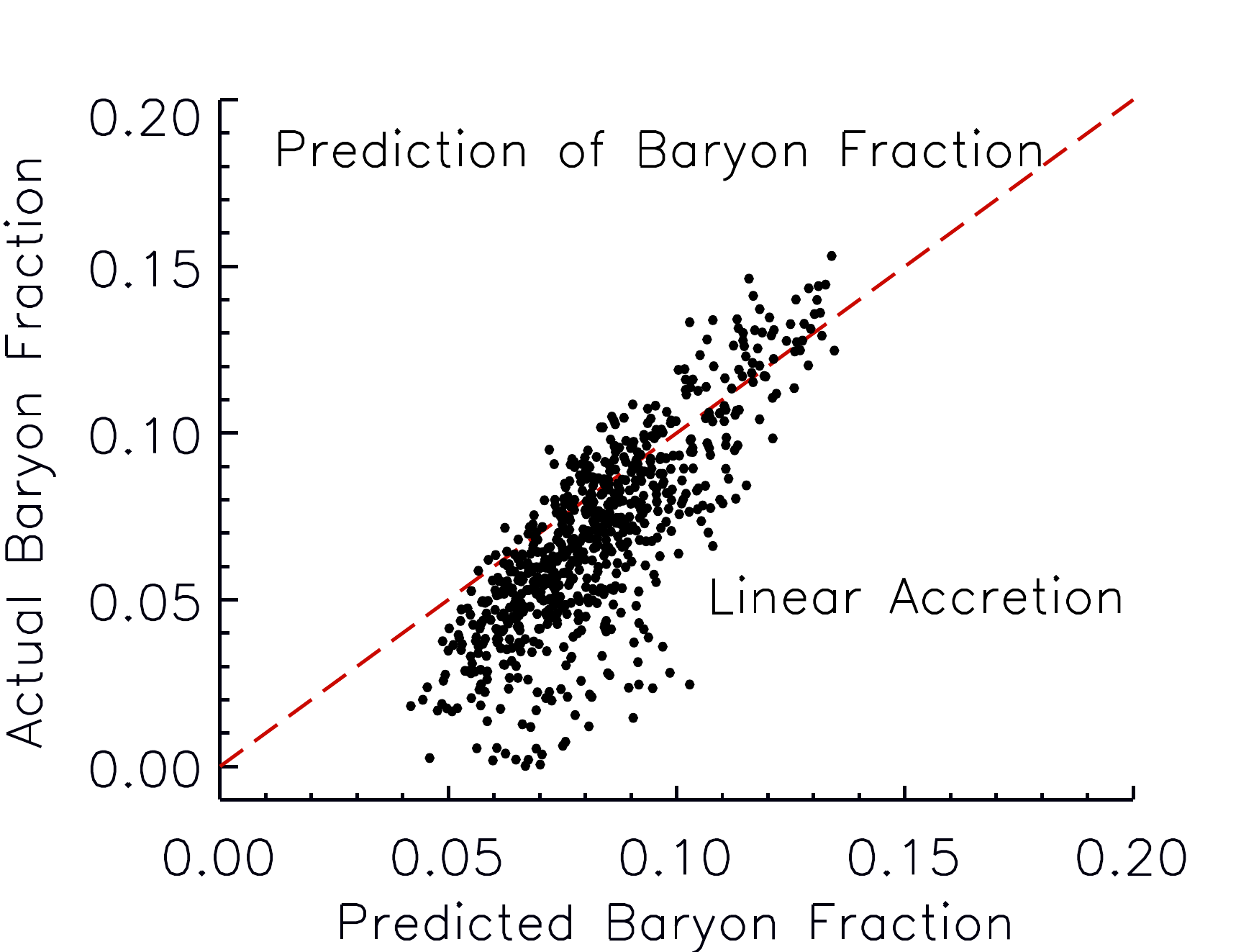}
    \end{center}
    \caption[1]{
      \label{bfractpredict}
      \bold{Predicting baryon fraction assuming filament
        accretion.
      }
      As described in Section~\ref{contrast}, the baryon fraction
      of the final collapse sphere for the dark matter
      is predicted from the baryon fraction
      at turnaround and the final collapse fractions of the dark
      matter and gas.  The computation assumes movement
      of the gas along a uniform filament.  Each point
      represents a single galaxy.  The abscissa shows the
      predicted value and the ordinate the actual one.
      The dashed, red line represents agreement of the
      prediction with the actual.   The $5.4$\% of
      galaxies for which the dark matter does not collapse
      faster than the gas are excluded as well as the galaxies
      where the gas, unlike the dark matter, does not turn
      around at all.
    }
    \end{figure}
}

\newcommand{\cmdlimitsdarkcoll}{
   \begin{figure}
     \begin{center}
     \leavevmode
    \includegraphics[scale=0.5]{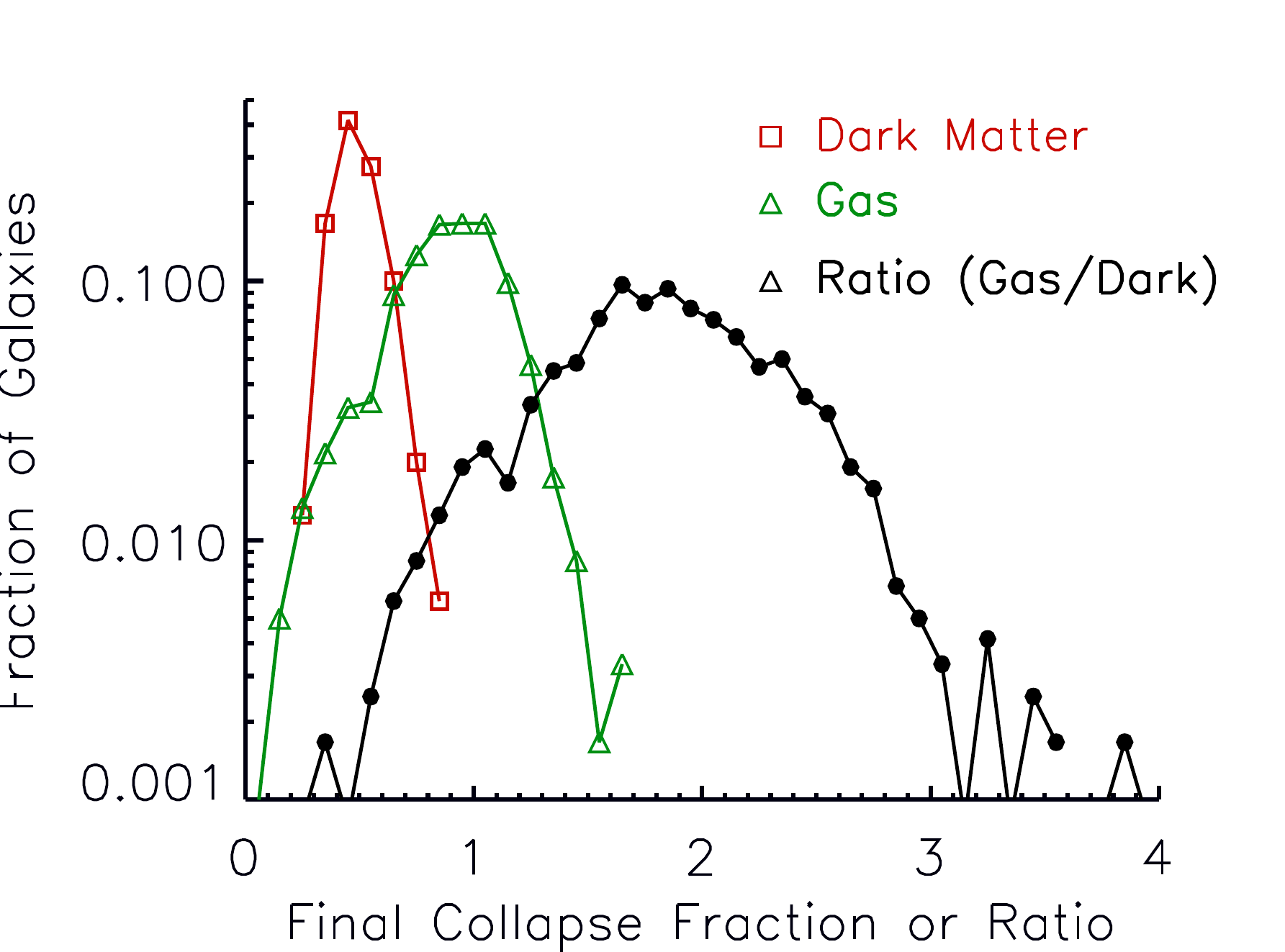}
    \end{center}
    \caption[1]{
      \label{limitsdarkcoll}
      \bold{Comparison of the movement of dark matter and
        gas into the galaxy.
      }
      Histograms show the final collapse fractions of dark matter
      (red squares)
      and gas (green triangles), and
      the ratio of the two for individual
      galaxies (black, filled circles).
      The graphs show that for $94.6$\% of the
      galaxies the dark matter collapses more rapidly than the
      gas.  The final collapse fraction is defined in
      Section~\ref{contrast}.
    }
   \end{figure}
}

\newcommand{\cmdturnavail}{
    \begin{figure}
    \begin{center}
      \leavevmode
    \includegraphics[scale=0.45]{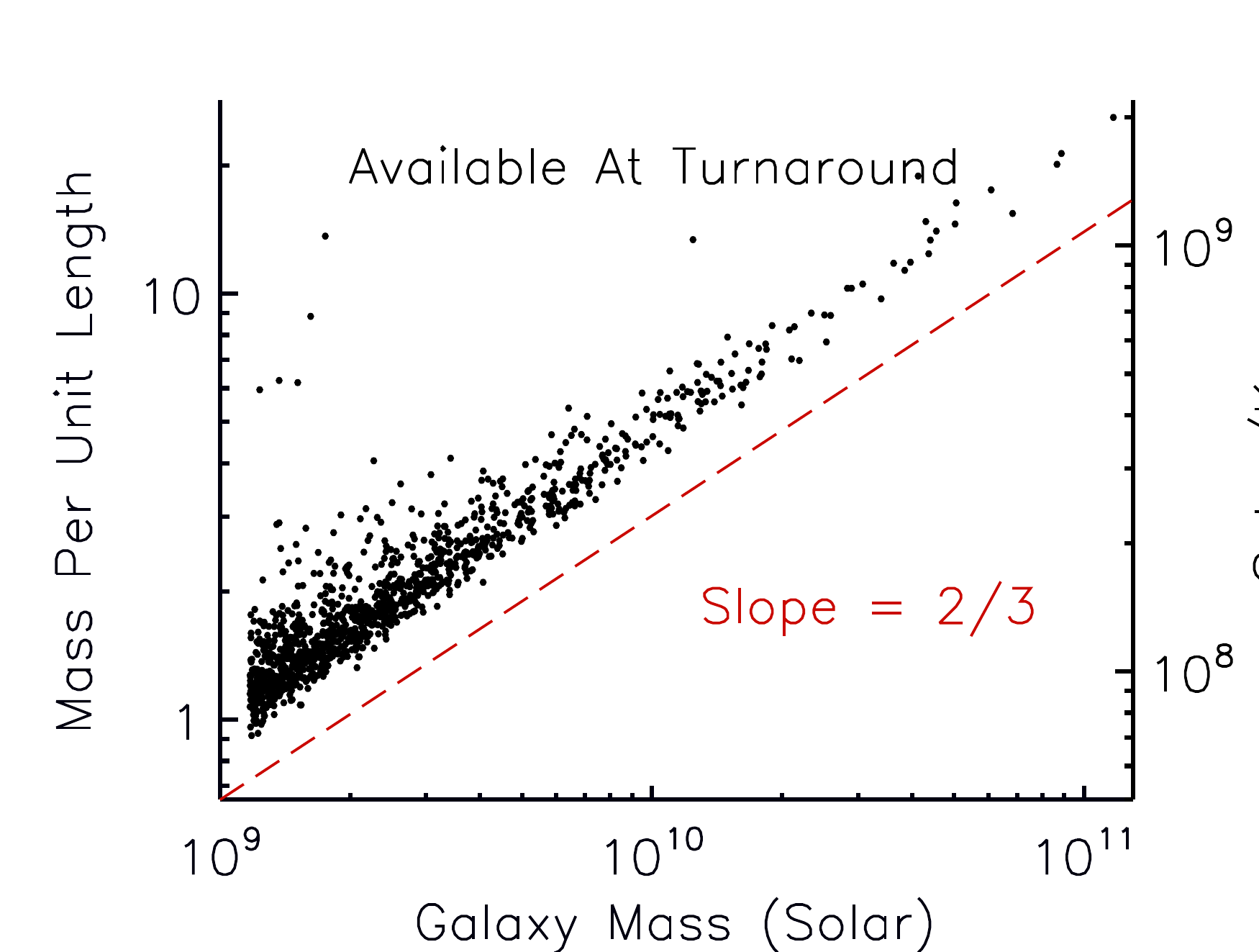}
    \end{center}
    \caption[1]{
      \label{turnavail}
      \bold{Available mass per unit length at turnaround as a
        function of the total mass of the galaxy.
      }
      Each symbol represents a single galaxy.  The available mass
      pre unit length is considered to be the ratio of the 
      total mass within
      the turnaround radius to the diameter of the turnaround
      sphere.  The left ordinate shows mass per unit length in
      the units used in Figure~\ref{figcorr},
      Figure~\ref{figcorrb}, and Figure~\ref{tbtfcssq} to compare
      the mass per unit length to the sound speed.
      The right ordinate shows the mass per unit length
      in units of solar/kpc for comparison.
      If the collapse is self-similar
      the expected slope of this log-log plot is two-thirds.  
    }
    \end{figure}
}

\newcommand{\cmdtbtf}{
    \begin{figure}
    \begin{center}
      \leavevmode
    \includegraphics[scale=0.5]{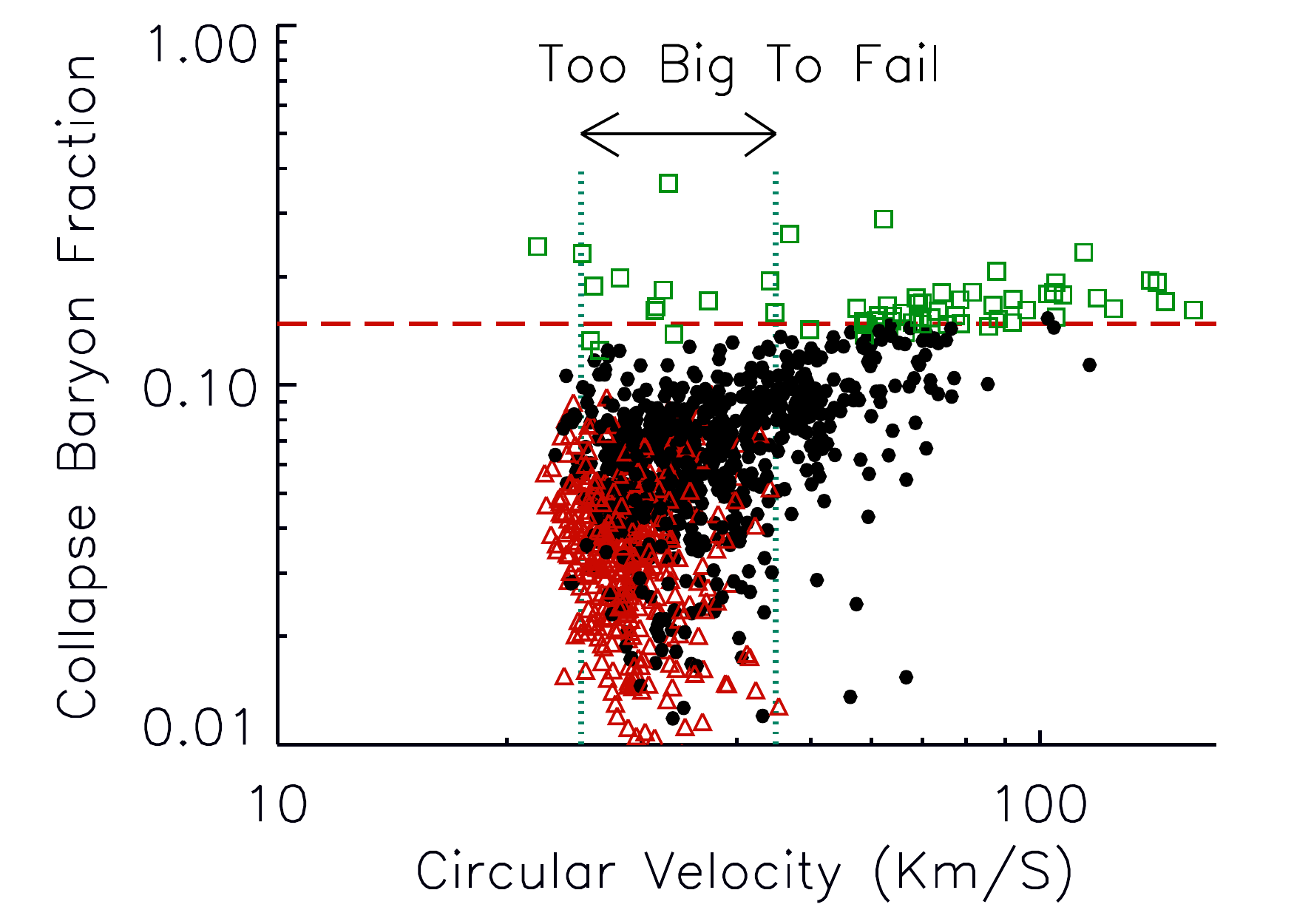}
    \end{center}
    \caption[1]{
      \label{tbtf}
      \bold{Scatter plot of the final baryon fraction of galaxies
        as a function of the halo circular velocity. }
      The ``too big to fail'' galaxies
      are considered to be those with circular velocity between
      $25$ and $45 \unit{km} \unit{s}^{-1}$
      \citep{papastergis16} as indicated by the
      blue, dotted, vertical lines.
      Shown on the ordinate is the baryon fraction of the
      final dark matter collapse sphere (see Section~\ref{contrast}
      for definition).        The red, dashed line
      indicates the cosmic baryon fraction.
      The circular velocity on the abscissa
      is computed from the dark matter in this sphere using
      Equation~\ref{vcircequation}.  The green squares represent
      the plunging-gas galaxies, namely those
      where the gas collapses as fast or faster than
      the dark matter.  The red triangles represent 
      the retreating-gas galaxies, namely those where
      the gas fails to turn around unlike the dark matter.  The
      black, filled circles are the galaxies between these two
      extremes, which we have called lingering-gas galaxies.
      These are the ones we suggest are the ``too big to fail''
      galaxies
      and which show the best agreement with
      an isothermal cylinder.  They are also the ones used to
      produce Figure~\ref{bfractpredict} and 
      Figure~\ref{bfractpredictcube}, which show that
      the baryon fraction can be predicted from the
      assumption that the matter collapse occurs along a filament
      rather than spherically.   The reduction of the baryon
      fraction below the cosmic value for these galaxies might
      make them invisible.
    }
    \end{figure}
}

\newcommand{\cmdtbtfcssq}{
    \begin{figure}
    \begin{center}
      \leavevmode
    \includegraphics[scale=0.5]{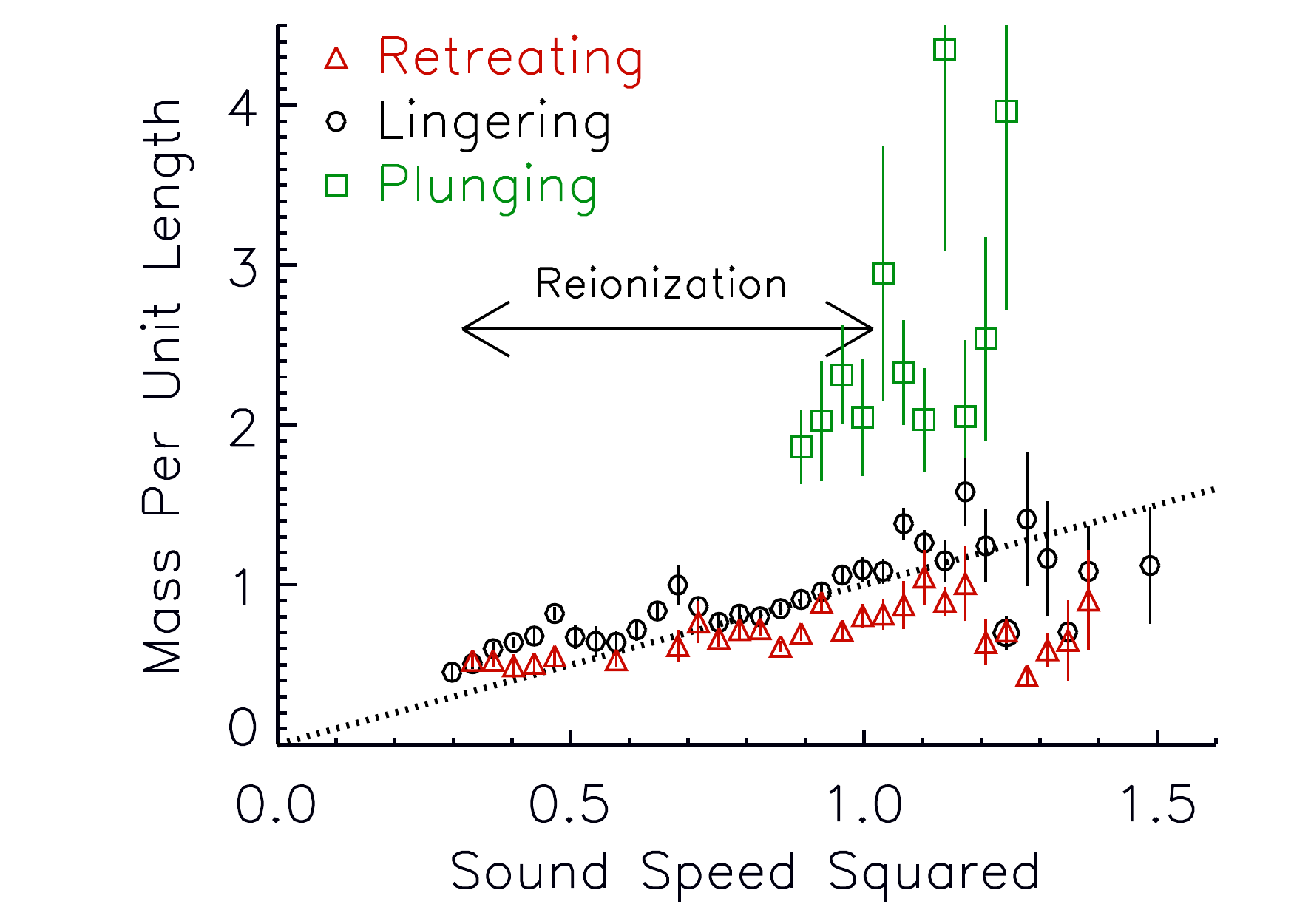}
    \end{center}
    \caption[1]{
      \label{tbtfcssq}
      \bold{Total mass per unit length as a function of sound
      speed.}
    The symbols show the average total mass (dark matter plus
    gas) for filaments associated with galaxies of the three
    types:  black open circles for lingering-gas, green
    squares for plunging-gas, and red triangles for retreating-gas.
    As for Figure~\ref{figcorr} and Figure~\ref{figcorrb}
    the units on the axes have been chosen to illustrate the
    expectation of proportionality for the model, shown as the
    diagonal, dotted line.
    Only bins having at least five members are shown for each
    type.  Vertical bars are single standard errors.  The range
    of sound speeds during the process of reionization is
    indicated by the black arrow.  
    }
    \end{figure}
}

\newcommand{\cmdexcessgas}{
    \begin{figure}
    \begin{center}
      \leavevmode
    \includegraphics[scale=0.5]{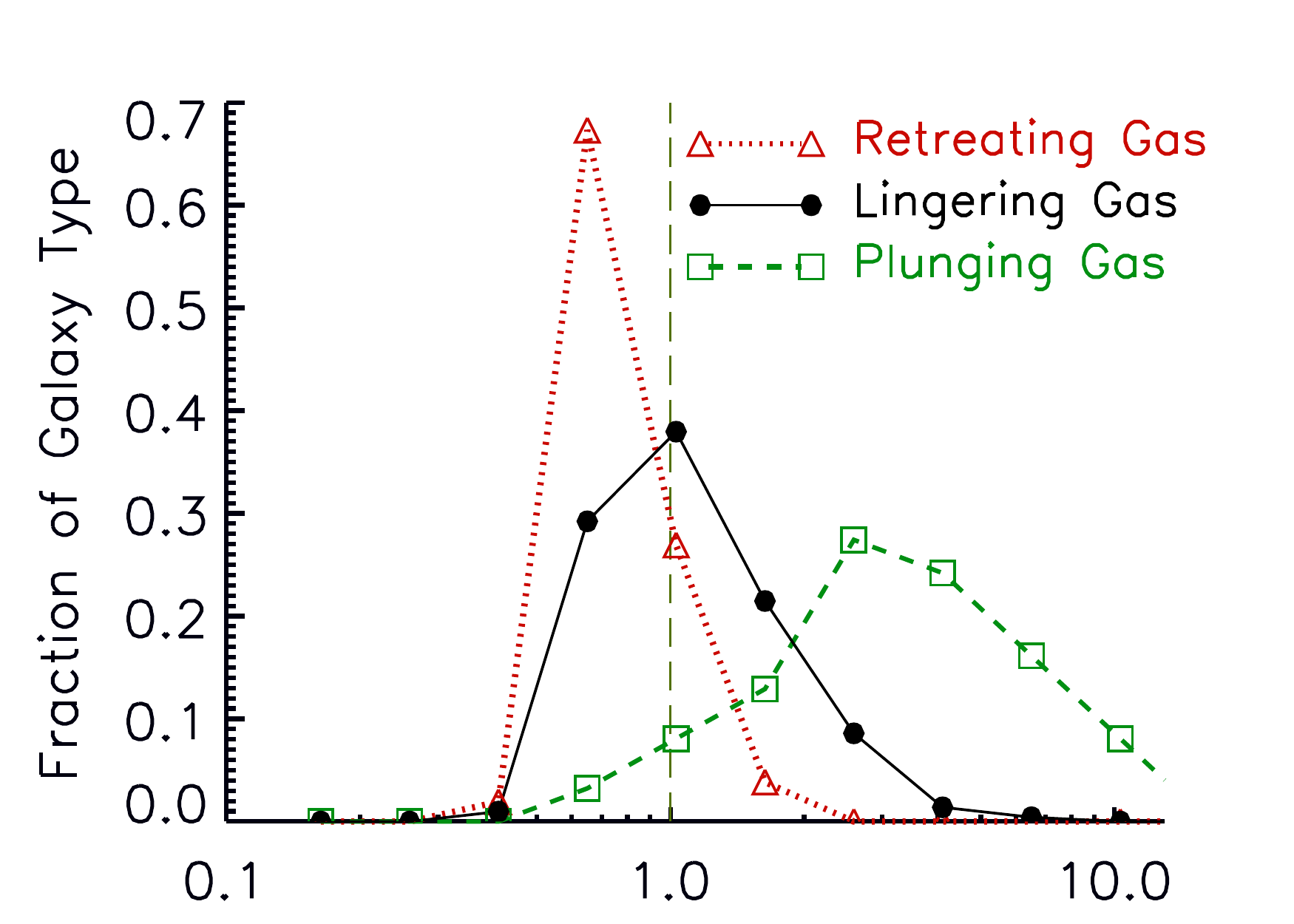}
    \end{center}
    \caption[1]{
      \label{excessgas}
      \bold{Available mass per unit length for the three galaxy
        types.}
      Histograms of the available mass per unit length for the
      three galaxy types.  The abscissa is the available mass per
      unit length in 
      the units used in Figure~\ref{figcorr},
      Figure~\ref{figcorrb}, and Figure~\ref{tbtfcssq} to compare
      the mass per unit length to the sound speed.
      The ordinate is the 
      fraction of that galaxy type.  The black, solid line with filled
      circles represents lingering-gas galaxies.
      The green, dashed line with squares represents plunging-gas
      galaxies.  The red, dotted line with triangles represents
      the retreating-gas galaxies.  The vertical, dark green,
      dashed line at $1.0$
      is the mass per unit length 
      corresponding to one unit in the figures showing the
      correlation with the square of the sound speed.
    }
    \end{figure}
}

\newcommand{\cmdbaryloss}{
    \begin{figure}
    \begin{center}
      \leavevmode
    \includegraphics[scale=0.5]{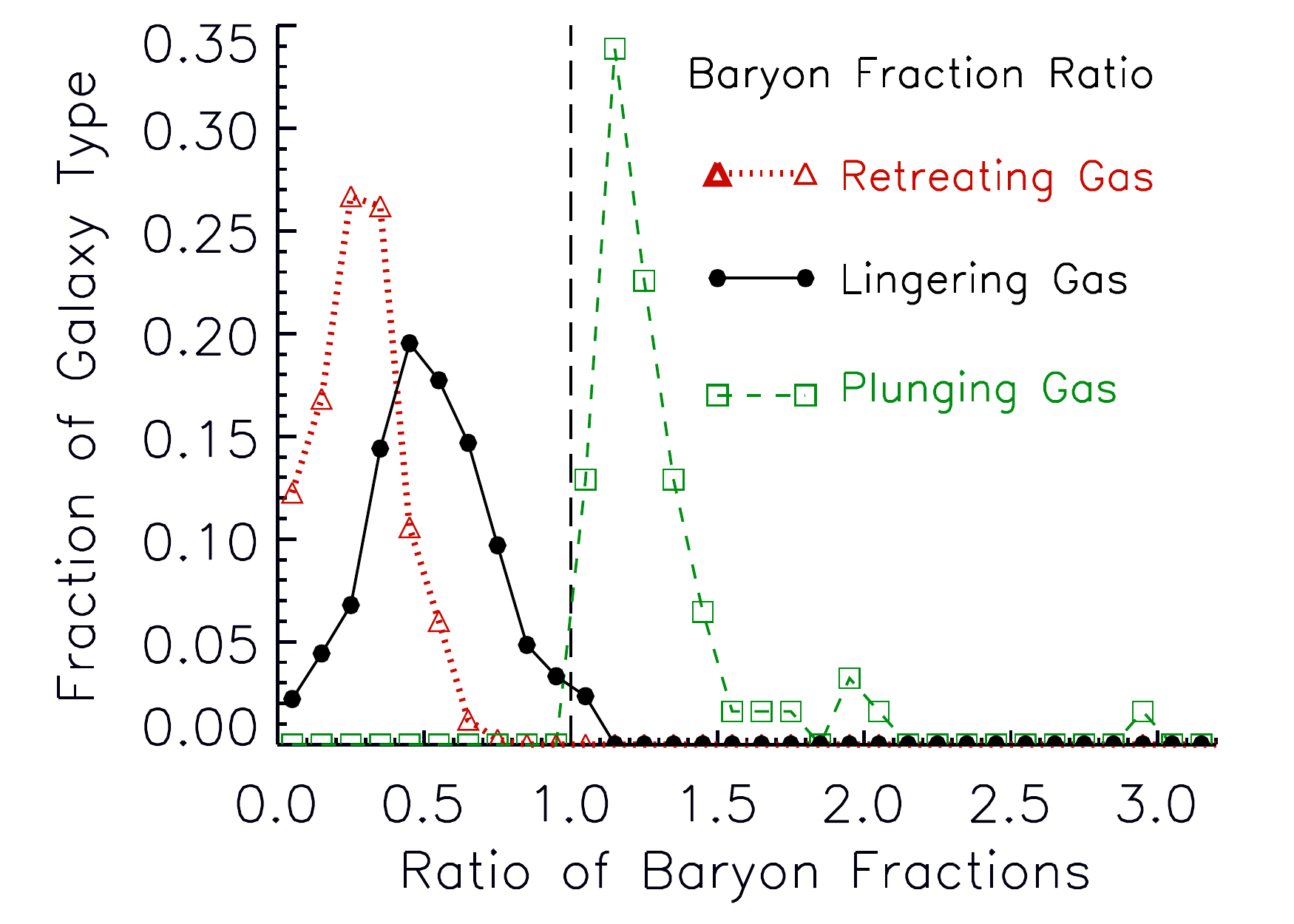}
    \end{center}
    \caption[1]{
      \label{baryloss}
      \bold{Change in baryon fraction after turnaround.}
      The abscissa of these histogram shows the ratio of the 
      baryon fraction of the final collapse sphere 
      to that of the turnaround sphere.  The ordinate shows
      the fraction of galaxies of each of the three types.  The
      vertical, black, dashed line indicates no change during
      this period.
    }
    \end{figure}
}

\newcommand{\cmdtally} {
    \begin{figure}
    \begin{center}
      \leavevmode
\includegraphics[scale=0.5]{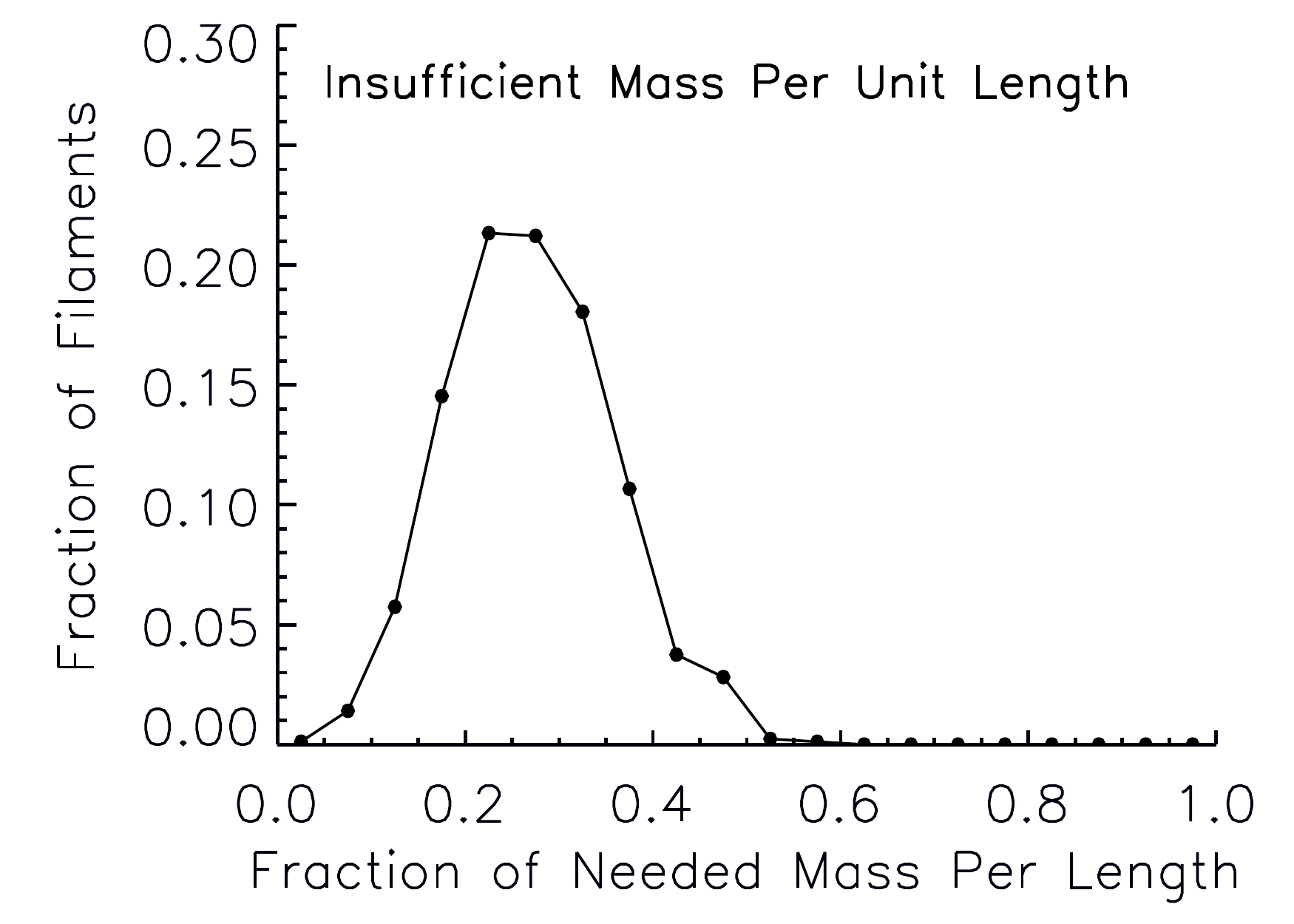}
    \end{center}
    \caption[1]{
      \label{tally}
      \bold{Virial budget.}
      This figure compares the mass per unit length of a
      filament to that required to balance in a virial
      equilibrium  both the sound
      speed of the gas and the bulk velocities of the dark
      matter and gas as explained in 
      Section~\ref{sectiondarkmodel}.
    }
    \end{figure}
}

\newcommand{\cmdvprof}{
    \begin{figure}
    \begin{center}
      \leavevmode
\includegraphics[scale=0.5]{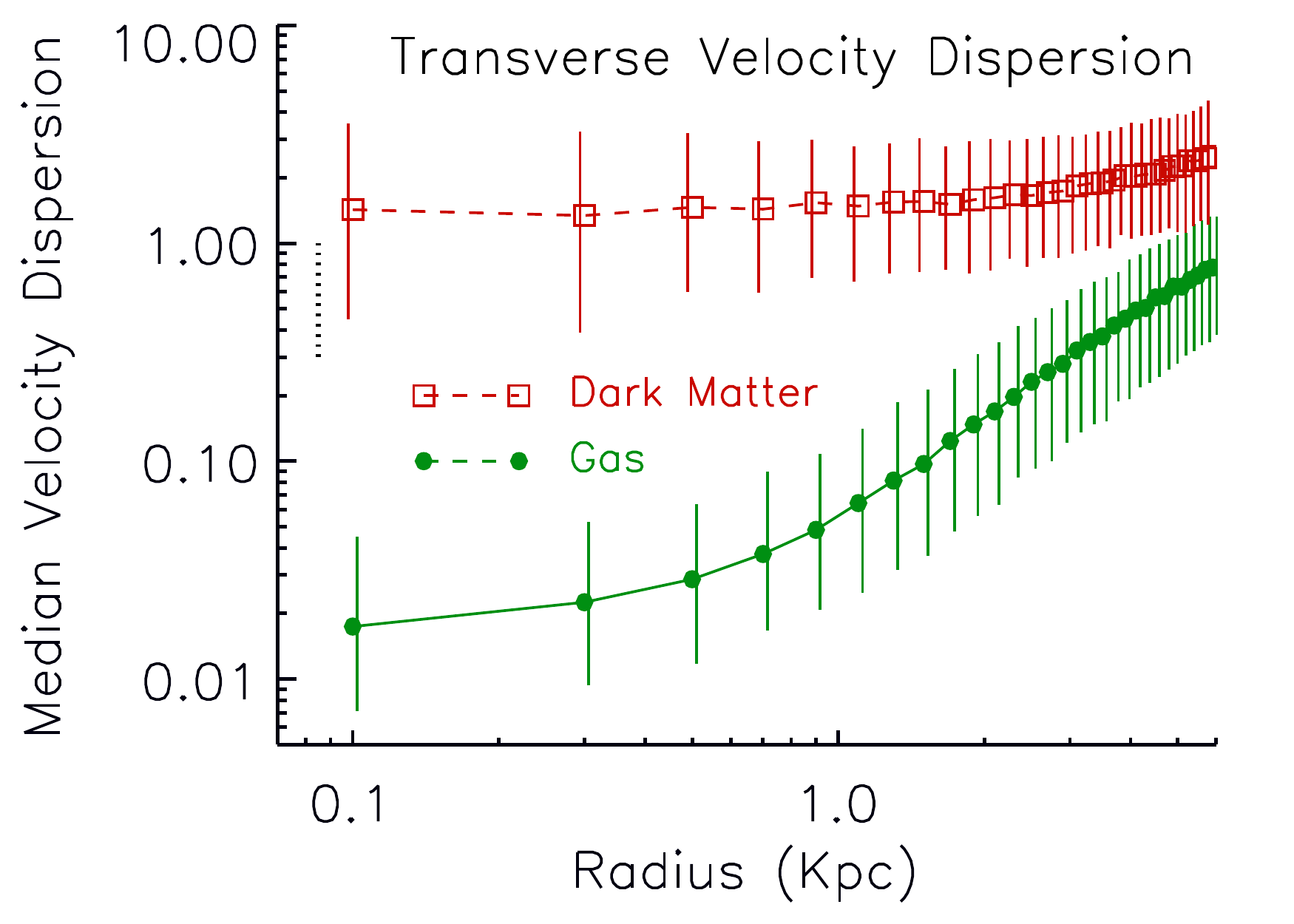}
    \end{center}
    \caption[1]{
      \label{vprof}
      \bold{Transverse velocity dispersion profiles.}
      The median transverse velocity dispersion of the 
      dark matter and gas
      is plotted as a function of distance from the axis of the
      cylinder.  Only filaments from lingering-gas
      galaxies that have an average sound speed squared
      of less than one are included.  Dark matter is 
      represented by  the dashed, red line with squares and
      gas by a solid, green line with
      filled circles.  The vertical lines at each point on the
      curves indicate the range of twp-thirds of the values.  
      The vertical, dotted, black line at the left 
      shows the range
      of the sound speed squared of the filaments.in our
      usual units (see Figure~\ref{figcorr}.  The units for
      the sound speed squared are chosen to correspond to the
      expected mass per unit length
      (see Figure~\ref{figcorr} for a description of these units).
      The units for the bulk velocities squared of the 
      gas and dark matter
      are chosen to be twice 
      that for the sound speed in order to account for the
      difference between Equation~\ref{isoequation} and
      Equation~\ref{velequation}.  The unit corresponds to
      a velocity of $25.7 \unit{km} \unit{s}^{-1}$.
    }
    \end{figure}
}

\begin{document}

\maketitle

\begin{abstract}
We propose a physically based, analytic model for
intergalactic filaments during the first gigayear of
the universe.  The structure of a filament is based
upon a gravitationally bound, isothermal cylinder of gas.
The model successfully predicts for
a cosmological simulation the
total mass per unit length of a filament (dark matter plus gas)
based solely upon the sound speed of the gas component, 
contrary to  the expectation for collisionless
dark matter aggregation.  In the model, the
gas, through its hydrodynamic properties, plays a key
role in filament structure rather than being
a passive passenger in a preformed dark matter
potential.  
The dark matter of a galaxy follows the classic equation of
collapse of a spherically symmetric overdensity
in an expanding universe.  In contrast, the gas usually
collapses more slowly.  The relative rates
of collapse of these two components for
individual galaxies can explain the varying baryon
deficits of the galaxies under the assumption
that matter moves along a single filament passing
through the galaxy centre, rather than by spherical
accretion.  The difference in behaviour of the dark 
matter and gas can be simply and plausibly related to
the model.  
The  range of galaxies studied includes that of the
so-called ``too big to fail'' galaxies, which are thought
to be problematic for the standard $\Lambda$CDM
model of the universe.   The isothermal-cylinder model suggests
a simple explanation for why these galaxies are, unaccountably,
missing from the night sky.
\end{abstract}


\section{Introduction}
\label{intro}
In the currently popular $\Lambda$CDM model of the universe,
collisionless dark matter dominates over baryons by a factor
of nearly six (see \citealt{frieman08} for review).
A two-stage model for galaxy formation \citep{white78}
has survived as a general paradigm, in which
the overall architecture of the universe is
formed by the potential wells formed by the collisionless
gravitation of the dark matter For reviews see \citet{frenk12,kravtsov12,
conselice14,somerville15}.
The luminous structures that might be observed then would
result from dissipational processes of the
baryons trapped within these wells.
This paradigm has justified the extensive use of N-body
simulations using only dark matter to understand structure
formation in the universe

The present paper challenges this paradigm in the case
of intergalactic filaments during the first gigayear.
Using a cosmological simulation
that includes gas hydrodynamics, radiative transfer, and chemistry
in addition to dark matter, we show that
the mass per unit length of intergalactic
filaments depends upon the sound speed of the gas
in the manner expected if the filaments were
gravitationally bound, isothermal cylinders of gas
\citep{stod63,ostriker64} with
dark matter mixed in.

An important implication of this model is that the
filaments should have a preferred size
as deduced from the simple analytic expression for their
structure.  This is because
the mass per unit length of such a cylinder depends
solely upon the temperature and ionization state of the
gas, which in turn are constrained by the Lyman alpha
cooling floor.  A preferred size for intergalactic filaments
is not expected from
collisionless structure formation by dark matter 
alone\footnote{Dark matter does have a preferred 
length scale arising from the
sound horizon at recombination.  However, this scale is vastly
greater than that of  galaxies and their attached filaments.}.

\cmdbaryrich

The present study was motivated by our previous simulation studies
\citep{harford08,harford11}, which showed that gas and dark matter
may assume very different structures
shortly after the end of the first gigayear
at a redshift of $5.134$.
For example, Figure~\ref{baryrich}  shows
separate images of gas and dark matter for the same
projection of a sphere centred upon a galaxy.
The gas appears as a relatively smooth filament  while the
dark matter is more irregular, with discrete clumps positioned along
the filament.  We showed that the
gravitational potential of the gas resembles that of
a gravitationally bound, isothermal cylinder in cases
where the filament is highly enriched in baryons.

In investigations that led to the present paper,
we began to explore the development of the filaments throughout
the first gigayear of the simulation,
in order to understand the origin of the isothermal structure.
We found that baryon enrichment of filaments emerges late in the first
gigayear.  Earlier most of the filaments are not enriched,
and yet, as we show in this paper, can be described as
gravitationally bound, isothermal cylinders, provided that we
make the simplifying assumption that the dark matter
adds to the gravitational potential as if it were
uniformly mixed with the gas.

Current thinking ascribes an important role to
intergalactic filaments in the transport of gas
into the small galaxies that form at high redshift.
Two modes of gas entry into galaxies have been
described in the literature.  In the ``hot mode''
incoming gas is shock heated when it encounters
the potential well of the halo \citep{rees77}.
The shocked gas
is elevated in temperature to where it can cool
efficiently
and then enter the halo.  This mode is thought
to predominate
for galaxies larger than a few times
$10^{11}\unit{M}_\odot$ and at low redshifts.
It is thought that for smaller galaxies,
like the ones we have studied,
the gas is not heated because the smaller
potential wells cannot sustain shocks
\citep{birnboim03}.
In this ``cold mode'' scenario, gas passes directly into
the galaxy without supersonic heating,
perhaps mediated by intergalactic
filaments \citep{birnboim03,binney04,katz03,keres05,
dekel06,ocvirk08,keres09b,dekel09d,brooks09}.

A gravitationally bound, isothermal cylinder provides
a physical framework for thinking about the motion of
matter along an intergalactic filament.
In the present paper, we show that the gas is generally retarded
in its entry into the galaxy centre relative to the dark matter.
We suggest that it is the hydrodynamic pressure of the gas,
counter-balancing the inward force of gravity, that causes
 this behaviour.

We show that the different behaviours of dark matter and gas
are consistent with a simple scheme for galaxy formation
in which the dark matter collapses according to collisionless theory,
while the gas tends to remain in extended, isothermal cylinders
in which infall into the galaxy is retarded by the pressure of the gas.

In addition we show that the different rates of accretion of 
dark matter and gas predict roughly  the
baryon deficits of the final galaxies under the assumption
that matter moves along a single filament passing
through the galaxy centre rather than by spherical
accretion.

Understanding the structure of intergalactic filaments
may not only provide insights into the
mechanisms and pace of early galaxy formation,
but may also place limits upon the sizes of galaxies
that can form from a filamentary precursor.
In this way the model might be
relevant to the so-called ``missing satellite'' problem
\citep{kauffmann93,klypin99,moore99},
which refers to the discrepancy
between the observed number of satellite galaxies
of the Milky Way and the predicted number of
dark matter haloes from N-body simulations containing
only dark matter.  An additional, ``too big to fail'' problem
has also arisen in which a class of dark matter 
haloes thought to be
massive enough to support star formation fail to be observed
\citep{boylankolchin09}.
The model suggests a simple explanation for this anomaly.
Resolution of these
problems is critical to the viability of the
cold dark matter hypothesis.

Most previous studies on intergalactic filaments
have dealt with larger
structures at later times in a more complex universe.
Our study suggests that the intergalactic filaments
associated with small galaxies during the first
gigayear, and present during the critical period of
reionization, may be a
relatively simple model system to study.

The plan of the paper is as follows.  In Section~\ref{methods}
we describe the self-consistent simulation of the first gigayear
that we have analyzed.  Section~\ref{model} gives an overview of
principal features of the model with illustrative diagrams.
Section~\ref{results} compares the model to the simulation.
Section~\ref{galdefinition} sets out exactly what we mean by a
galaxy in the simulation and how we decide where it is at any
given time.  

Section~\ref{filstructure}
begins with a brief summary of the algorithm
for identifying intergalactic filaments associated with specific
galaxies.  The algorithm is described in more detail with
diagrams in the Appendix.  Then the basic structural equations
of a gravitationally bound, isothermal cylinder are laid out
as developed by \citet{stod63} and \citet{ostriker64}.
The agreement of the filaments with such structures
is then tested by comparing the total mass per unit length to that
predicted by the model.   

Section~\ref{darkmotion} shows that the collapse
of the dark matter can be described by the textbook collapse
of a spherically symmetric overdensity.  
Section~\ref{contrast}
describes how we compare and contrast the collapse of gas
and dark matter onto the final galaxy. Three categories of galaxies
are distinguished in this regard:  plunging-gas, retreating-gas,
and lingering-gas.
Section~\ref{baryondeficits}
relates the findings to the baryon deficits of the final
galaxies. The rates of movement of the gas and dark matter
in the lingering-gas galaxies
argue for collapse along a single filament as opposed to
spherical accretion.  

Section~\ref{retard} relates the three
categories of galaxies to the formation of isothermal cylinders
and to the mass available to form filaments.
Section~\ref{sectiontbtf} relates the
model to the so-called ``too big to fail'' galaxies.

Section~\ref{sectiondarkmodel} explores an 
alternative model in which 
filament structure is determined primarily by 
dark matter.

Section~\ref{sectionalign}
presents evidence that the filaments are aligned
as predicted by the model.

Finally the results are discussed and summarized in
Section~\ref{discussion} and Section~\ref{sectionsummary} respectively.



\section{Cosmological simulation}
\label{methods}

As in our previous work
\citep{harford08,harford11},
the results reported in the present paper
are based on a flat $\Lambda$CDM cosmological simulation
that includes gas hydrodynamics, radiative transfer, and chemistry
in addition to dark matter.
The simulation followed a $8 h^{-1}$ comoving Mpc cube
on a $256^3$ grid
up to a redshift of $5.13$ using a
``softened Lagrangian hydrodynamics'' (SLH-P$^{3}$M) code
\citep{gnedin95,gnedin_bertschinger_96}.
The cosmological parameters are  $\Omega_{m} = 0.27$, $\Omega_{b} = 0.04$,
$\sigma_{8} = 0.91$, and $h = 0.71$.

Dark matter is computed with collisionless particles of mass
$2.73\times10^{6}\unit{M}_\odot$.
Gas dynamics is computed on a quasi-Lagrangian mesh
that deforms adaptively to
provide higher resolution in higher density regions.
The mass of a gas particle is initially
$4.75\times10^{5}\unit{M}_\odot$,
but each gas particle mass may adjust slightly  as the 
hydrodynamic computation evolves.

\section{Overview of Model}
\label{model}

A simple paradigm for gravitational collapse begins with
collapse in one dimension to produce a plane.
Further collapse in a second dimension produces a rod, which
then collapses in the third dimension to a quasi-spherical ball
\citep{zeldovich70,park90,bertschinger91,cen93}.
Visual inspection of the simulation
suggested that this paradigm would be a good starting point. 

\cmdmodeldiagram
\cmdmodelexample

It is common to assume that galaxies form at the intersections of
filaments.  This is  because large-scale views of simulations
at late times show galaxies
as nodes in a complex network of filamentous material.
This paper takes a different point of view for the
first gigayear.
We consider an individual filament to be an intermediate 
structure in the formation of a galaxy.  In the model
an individual galaxy can be traced back in time to a 
single rod,
in the centre of which a quasi-spherical galaxy will form 
as the contents of the rod collapse.
After the first gigayear, multiple collisions may lead to the build up
of complex networks, which we will not consider here.

In the model the rod is a
gravitationally bound, isothermal cylinder of gas with comingled
dark matter.  Initially the gas and dark matter are uniformly
mixed.  The dark matter then proceeds to collapse toward 
the centre of
the rod, where the future galaxy will be found.
The gas, in general, collapses more slowly than the dark matter
because the pressure of the gas retards its flow into the galaxy.

The rod, in its initial form and during differential
collapse, is shown schematically in Figure~\ref{modeldiagram}.
\boldemp{In the upper image the red, filled circles representing the 
dark matter are uniformly
mixed with the green squares representing the gas.
In the lower image, showing a later time, the dark matter
particles have begun to coalesce into the centre of the rod
leaving behind the gas particles.}
An actual example from the
simulation is shown in Figure~\ref{modelexample}.

\section{Comparison of the model to the simulation}
\label{results}

The purpose of this section is to compare the model proposed
in \S\ref{model}
to the cosmological simulation described in \S\ref{methods}.

\subsection{Galaxies}
\label{galdefinition}

The objects referred to as ``galaxies'' in the present paper
were identified at the end of the simulation,
at a redshift of 5.134 (1.15 gigayears after the big bang),
from the total mass distribution
by DENMAX \citep{bertschinger91}.
The positions of the individual dark matter
particles associated with each of these galaxies can be
followed throughout the simulation.
For a given galaxy defined by
DENMAX at the end
of the simulation, we define it
at earlier times as the arrangement of its
constituent dark
matter particles at that time.  The centre of the
galaxy at any time is defined as the centre of mass
of its dark matter particles,
and the formation of each galaxy is
followed with time in a local coordinate system relative
to that centre of mass.  In this coordinate system the dark matter
particles of each galaxy exhibit an initial expansion followed by
a well defined turnaround.
The transient, quasi-spherical sub-structures seen at intermediate 
redshifts are not
treated as separate galaxies. 

In order to
follow the development of single galaxies in time since the
big bang, we focus on just the $1{,}200$ largest galaxies
identified at the end of the simulation just after the first gigayear.
The galaxies considered range in total mass from
about $10^{9}\unit{M}_\odot$ to several
times $10^{11}\unit{M}_\odot$.
The total mass of the galaxy is defined to
include dark matter, gas, and stellar material.

\subsection{Isothermal Filaments}
\label{filstructure}

To avoid subjective bias in identifying filaments,
we adopt an objective algorithm
which is detailed in Appendix~\ref{filamentid}.  Shortly after
turnaround of the dark matter, the single rod of the model appears embedded in a
planar slab of material.  We take advantage of this situation to
select the filaments in the two dimensions of the principal
plane.  The filaments are selected
using only the gas component because its structure is more
regular than that of the dark matter.  Although the model in its
purest form is best understood as the collapse of a single rod,
we will refer to the two halves of the rod as separate filaments
because both halves are not always present in the structures
surviving our selection algorithm.  No more than two
filaments are selected for a single galaxy.  When two are present,
the filaments are usually oriented end-to-end as if part of a
single, continuous structure.

The mass per unit length, $\Upsilon$, predicted for a gravitationally bound,
isothermal gas cylinder is 
\begin{equation}
\label{isoequation}
\Upsilon = \frac{2 c_s^{2}}{G}
\end{equation}
where $c_s$ is the isothermal sound speed and $G$ is the gravitational
constant \citep{stod63, ostriker64}.
The sound speed $c_s$ at temperature $T$ is
\begin{equation}
\label{soundspeedequation}
c_s = \sqrtsign{\frac{kT}{1.22\mu m_H}}
\end{equation}
where $k$ is the Boltzmann constant, $m_H$ the mass
of the hydrogen atom and $\mu$ a mean particle mass to
take into account the ionization of the hydrogen.
The factor of $1.22$ corrects for the
contribution of neutral helium to the mean atomic weight.

The important point here is that the mass per unit length of a filament
depends only upon its temperature and ionization state, and
not upon the concentration of the gas in the
transverse direction,

The presence of dark matter adds gravitational field without any
additional pressure.  The simplest way, and the way we adopt, 
to account for the effect
of dark matter is to assume that the
gas and dark matter are uniformly mixed.  The result of this
assumption is that Equation~\ref{isoequation} predicts, not
the mass per unit length of the gas, but the total mass per
unit length including the dark matter\footnote{The amount of
stellar material in the filaments in negligible, and we neglect
it.}.  It is interesting to note that this scheme
means that if the relative amounts of gas and
dark matter in the filament vary, then they
must vary in opposite directions to keep the total in line
with the sound speed.  
This situation contrasts with the conventional assumption that
gas follows dark matter.

\boldemp{Appendix A describes the algorithm for
selecting regions of the filaments to study.
For detailed analysis, a filament segment is defined as
the gas and dark matter within a cylinder of length 
twelve proper kiloparsecs 
and a radius of six proper kiloparsecs centred on
the filament region selected as described in Appendix A.}
Even though the mass of the cylinder extends to infinity in
theory, we find that most of the mass
is within this proper radius.  The orientation of the axis of
the cylinder is determined by a principal component analysis.

We might expect the development of a filament to depend
upon the local thermal history.  Reionization occurs during
the first gigayear.  However, different regions of the
simulation reionize at different times.
For this reason
we present the structure of the
filaments using the square of the sound speed as a proxy for time.
This scheme accomodates the possibility that different
filaments are at different stages of development
at the same time.  The sound speed depends only upon the
temperature and ionization state of the gas and is independent
of the density.  It is computed as the \bold{average of the squares of
the sound speeds} of the individual gas particles in the filament
segment being analyzed.  The sound speed of a gas particle is
computed from its temperature and ionization state, which are
independently computed.  In the simulation the temperature and ionization are computed
self-consistently from the radiation produced at specific loci of star
formation.  There is no superimposed, ionizing field as in many
other simulations.

\cmdreionization

The course of reionization of the individual
filament segments is shown in
the scatter plot in Figure~\ref{triplet}, in which each point
is a filament segment at one of the six redshifts we studied.
The abcissa is the average of the square of the sound speeds
of the gas particles
in the segment.  The ordinate shows the mean mass per
particle (nuclei and electrons)
in the gas of the segment, which
decreases as a result of reionization.   It is clear
that the segments at any given redshift are present in a range
of different stages of reionization.
In this paper the square of the sound speed is always
expressed in the same units chosen 
to facilitate comparisons with the predictions
of the model in Figure~\ref{figcorr} and Figure~\ref{figcorrb}.  

\cmdfigcorr

Figure~\ref{figcorr} is at the crux of the argument for gravitationally
bound, isothermal cylinders.  It shows, over a range of sound speeds,
that the total mass per unit
length of the filament segments can be predicted from the sound
speed of the gas using Equation~\ref{soundspeedequation}.  The
value computed for the gas from this equation is divided by the
overall gas fraction of
the segment to give the total mass per unit length plotted in the
figure.  The units of the axes have been
chosen for convenience so that the model predicts a slope of one.
The ordinate shows
the average for filaments segments in bins of sound speed squared
on the
abscissa.  Bins having at least fifteen members, shown by
red squares, were chosen
for a linear regression analysis.  The slope obtained,
$1.04 \pm 0.12$, shows good agreement with prediction.

The prediction, shown by the dotted, diagonal line of unit slope
in Figure~\ref{figcorr} assumes for simplicity that the dark
matter is uniformly intermixed with the gas and contributes
to the gravitational field in proportion to its 
abundance\footnote{\boldemp{It is impractical to 
subdivide the filament
segments into regions with different amounts of dark
matter even though the dark matter is clumpy.  The 
gas tends to be smoother and not to clump with the dark
matter.}}
An alternative scheme, in which
the gas is concentrated in a relatively flat central part
of a dark matter potential well would predict
for most segments
a gas mass per unit length greater than observed
by a factor of about five,
since in the latter scheme the gas behaves as if
it were a pure gas cylinder without dark matter.
The measured mass per unit length is more consistent with
the model where the gas and dark matter are uniformly 
mixed.

In Figure~\ref{figcorr} the sparser bins to the right
(green triangles) show only a
general upward trend consistent with prediction.  Orange
circles indicate bins with a single member.

The period of best agreement
with the model corresponds to the period of
active reionization when the gas in the segments is partially
ionized.  The extent of this region is indicated by the horizontal,
two-headed arrow.  During this period most of the
change in sound speed is due to reionization.  The temperature
changes little.

One should not conclude from these results that the
model applies only
to the reionization period.
During the limited period of the
simulation only a few of the filaments have progressed
significantly beyond reionization,
and these are associated with the
largest galaxies.  We will argue in
Section~\ref{retard}  that 
excessive mass can
lead to deviation from the model.
However, the small numbers
of segments in these sparse bins make it difficult to tell just how
well the model works for the largest galaxies. 

\cmdfigcorrbden

Another variable that might be expected to affect the
filament structure is the proper density at turnaround.
Figure~\ref{figcorrbden} shows
that the total mass per unit length does indeed increase with
density at turnaround as one might expect.
To demonstrate that the correlation of Figure~\ref{figcorr}
is not an artifact resulting from a correlation of 
sound speed with density at turnaround, 
we tested filaments from a restricted range of 
turnaround densities.
Figure~\ref{figcorrb}
shows a version of Figure~\ref{figcorr}
in which only galaxies within a narrow range of turnaround densities,
those between the two vertical, dotted lines in 
Figure~\ref{figcorrbden}, are included.
Figures~\ref{figcorr} and~\ref{figcorrb} are almost identical.

\cmdfigcorrb

\subsection{Collapse of Dark Matter}
\label{darkmotion}

\cmdexcycloid
\cmdoverden

The simplest theory of galaxy formation in an expanding
universe is that of the collapse of a spherically symmetric overdensity
\citep{peacock99}.   \cite{sugerman00} find that,
despite its simplifying assumptions, the spherical model provides
reasonable predictions for properties of dark matter haloes.
The spherical model predicts that the radius $r$ of a mass $m$
evolves with time $t$ as a cycloid
\citep{peacock99},
\begin{equation}
\label{cycloideq}
r={r_{\rm turn} \over 2}(1-\cos\theta),
\quad
t=\sqrt{r_{\rm turn}^3 \over 8Gm}(\theta-\sin\theta).
\end{equation}
The cycloid describes expansion from zero to a maximum radius,
the turnaround radius $r_{\rm turn}$,
as $\theta$ goes from $0$ to $\pi$,
followed by contraction
as $\theta$ goes from $\pi$ to $2\pi$.
To compare this theory with a given galaxy in the simulation, 
the average radius of the dark
matter particles is computed at a series of times before 
and after turnaround.
To compare these radii to the cycloid prediction, the mass $m$ 
in equation~\ref{cycloideq} is 
computed from the actual turnaround sphere using
the theoretically expected overdensity of $5.55$ 
\citep{peacock99}.
 
We find the spherical model
to be an excellent starting point for understanding
the collapse of the dark matter of each galaxy.
Figure~\ref{excycloid} shows, for an example galaxy,
the average radius of the galaxy's dark matter particles
from their centre of gravity and the corresponding cycloid.
Virtually all the galaxies fit a cycloid at least up until
turnaround,

The fitted cycloid is translated
slightly in time so that the turnaround time coincides with that
in the simulation.
We find that the cycloid generally begins its ascent either at
the beginning of the simulation or slightly before.
We interpret this result to mean that the
galaxies originate from fluctuations
imposed at the start of the simulation,
rather than from density variations resulting from subsequent events.

Figure~\ref{overden} shows
histograms of the overdensity, time, and radius at turnaround.
The approximate agreement of the overdensity
with the theoretical value of $5.55$ \citep{peacock99}
supports our interpretation for the
dark matter collapse as the collapse of a 
spherically symmetric overdensity.

\subsection{Contrasting Motion of Dark Matter and Gas}
\label{contrast}

Unlike the dark matter particles, the individual gas particles
in the simulation cannot be traced throughout the simulation.
Rather, they are
generated anew after each hydrodynamic timestep.
Since the structures
are continually moving and changing shape, it is difficult
to pin down gas movements on a small scale.
What we can compute unambiguously, however, is an overall measure of
collapse.   We compute this separately for the gas and the
dark matter so that we can compare their relative rates of
motion into the centre of the galaxy.

\cmdsolo

At each redshift
we compute as a function of redshift
the proper radius of a sphere centred on the galaxy that includes
just that mass of gas or dark matter that was present
inside the turnaround radius at the time of turnaround
(the turnaround radius is the average radius of the dark matter particles
of a galaxy at the time of turnaround).

Figure~\ref{solo}
shows some individual galaxy histories that illustrate the
types of results obtained.  The abscissa shows the cosmic
scale factor starting from the time of turnaround,
which is slightly different for different galaxies.
The ordinate is the radius as a fraction of the turnaround radius.
Most of the time the gas collapses more slowly than the dark
matter as shown in the first two graphs.  In the left graph
the gas does not collapse fast enough to outrun the expansion
of the universe, and the gas radius increases with time.
We call this the ``retreating
gas'' type of galaxy.  We will see that these are among the
least massive of the galaxies we have studied.  In the
``lingering gas'' type, shown in the middle graph, the gas collapses but more slowly
than the dark matter.  We will see in later sections that
the filaments associated with these galaxies have the best
fit to an isothermal cylinder.  Finally, the right graph shows
the gas collapsing more rapidly than the dark matter,
the ``plunging gas'' case.  There are $416$ retreating-gas 
galaxies,  $722$ lingering-gas galaxies,
and $62$ plunging-gas galaxies.

We will refer to the sphere containing the turnaround amount of mass
as the \textit{collapse sphere} for the gas or dark matter as the
case may be for the time in question.
The proper radius of the collapse sphere
divided by the proper turnaround
radius we will call the \textit{collapse fraction}.  When the radius is
taken at the end of the simulation we will refer to this
ratio as the \textit{final collapse fraction} of the gas or dark matter.

\cmdlimitsdarkcoll

Figure~\ref{limitsdarkcoll} shows histograms of the final collapse
fractions of dark matter and gas for the $1{,}200$ galaxies
studied in this paper.
The histogram for the dark matter (the red curve with squares)
peaks at a collapse fraction of about one half,
in agreement with the virial expectation for
collisionless particles subject only to gravitation.
In contrast, the gas
(green curve with triangles)
generally collapses more slowly.

The black curve with filled circles in Figure~\ref{limitsdarkcoll}
shows a histogram of the
ratio of the collapse fraction of the gas to that of the
dark matter for the same galaxy.  The small fraction of the
area under the curve to the left of a ratio of $1.0$
shows that most of the time the gas collapses more slowly
than does the dark matter.
We suggest that the pressure in the gas filament counter-balancing
the force of gravity is responsible for this difference in behaviour.

\cmdtimescattgfract

The end of reionization  marks the beginning of a
period of increasing deviation of the gas fraction of the filament
from the cosmic value as shown in the scatter plot in
Figure~\ref{timescattgfract}.   Here each open circle represents
a filament segment.

\cmdcollapsegf

We suggest that this effect occurs when rapidly collapsing dark
matter effectively leaves behind the gas in the filaments.
Figure~\ref{collapsegf} shows the average gas fraction
of filament segments as a function of the collapse fraction
of the dark matter of the associated galaxy at the 
redshift of the filament.  As the dark matter of a galaxy
collapses, the filaments of that galaxy can become more
baryon rich.
Note that in this figure maximal
collapse is at the left and minimal collapse at the right.

\subsection{Filamentary as Opposed to Spherical Accretion.}
\label{baryondeficits}

Unlike many of the final galaxies, the
turnaround sphere has a baryon fraction close to
the cosmic value.
The differential collapse of gas and dark matter can
lead to significant baryon deficits in the final galaxies.
Figure~\ref{baryturncat}
compares histograms of the baryon fraction at turnaround
and in the final galaxies.

\cmdbaryturncat
\cmdbfractpredict
\cmdbfractpredictcube

The differing motions of the dark matter and gas
that we have
just described in Section~\ref{contrast}
can be related in a simple way to these
baryon deficits.  Consider,  for simplicity, the
lingering-gas situation in which the dark matter collapses faster
than the gas.  
At the end of the simulation
the final collapse sphere for the galaxy's
dark matter will have a baryon fraction less
than that of the turnaround sphere and closer 
to that of the baryon fraction of the final galaxy.
The gas in this sphere is some fraction of the gas at
turnaround.  If, for simplicity, we assume that the gas
in the both the initial turnaround sphere and the final
dark matter collapse sphere is in the
form of a uniform filament that runs along the
diameter of both spheres, then
the fraction of the turnaround gas that is in the
dark matter collapse sphere can be estimated from the ratio
of the radius of this sphere to that of the
larger gas collapse sphere.  Thus
the baryon fraction of the final dark matter
collapse sphere,
can then be computed knowing the baryon fraction at
turnaround.


Figure~\ref{bfractpredict} shows that this simple
filamentary scheme
predicts the baryon fraction of the final dark matter
collapse sphere quite well.  Each point on the graph
represents a single galaxy.  The abscissa is the
predicted baryon fraction and the
ordinate is the actual one.   The red, dashed
line represents agreement between the two.
Excluded from this
plot are plunging-gas galaxies where the gas collapses as fast or
faster than the dark matter.  These are mostly the
largest galaxies.  Also excluded are the retreating-gas 
galaxies where
the gas fails to turn around even though the dark
matter does.  These are the galaxies where the
final collapse fraction of the gas is greater than
or equal to one.

Figure~\ref{bfractpredictcube} show the contrasting
predictions if
the relevant ratio of radii were instead cubed, as would be more
appropriate for spherical accretion.  The filamentary
accretion model is a much better fit than is the spherical one.

The baryon fractions shown in Figure~\ref{bfractpredict}
and Figure~\ref{bfractpredictcube} are for the final dark matter
collapse sphere and are somewhat lower
than those of the final galaxies as identified by the
galaxy finding algorithm DENMAX.  Since the turnaround
sphere is based upon the average radius of the dark matter
particles at turnaround, we might expect the final dark
matter collapse
sphere to represent an inner sphere of the galaxy.
DENMAX  might be adding to this 
a more baryon rich region
at the periphery of the galaxy.

From the results of this section we suggest that a major
cause of baryon deficits in the galaxies we have studied
is the retarded motion of the gas relative to the dark
matter.  This situation is in constrast to a mechanism
whereby gas already in the halo is subsequently expelled.

\subsection{Isothermal Cylinders Retard Gas}
\label{retard}

In Section~\ref{contrast}
we distinguished three types of galaxies.  The ones where
the gas moves toward the galaxy as fast or faster than the dark matter
we referred to as ``plunging-gas'' galaxies.
Those where the gas shows no net movement toward the galaxy
we referred to as ``retreating-gas'' galaxies.
Those where the gas moves toward the galaxy but not as fast
as the dark
matter we referred to as ``lingering-gas'' galaxies.

The filaments of these three galaxy types
have been tested separately
for evidence of gravitationally bound, isothermal cylinders.  
Figure~\ref{tbtfcssq} shows the correlation of the mass per unit
length with the square of the sound speed.   The filaments from
the lingering-gas galaxies (black, open circles)
show the best match.  The filaments of the retreating-gas galaxies
(red triangles) fail to keep up with the expansion of the universe.
The plunging-gas galaxies, shown by the green squares have too
much mass for isothermal cylinders.

\cmdtbtfcssq
\cmdexcessgas

These contrasting results
can be understood in terms of the key property of a
gravitationally bound, isothermal cylinder that
the mass per unit length is limited by the 
sound speed of the
gas.  We might expect the cylinder to break down if the
overall collapsing mass is
overwhelmingly larger than can be accomodated at the
current sound speed.
This situation would correspond to the plunging-gas case.
On the other hand, the filaments of the retreating-gas galaxies
might have too little matter to
withstand the expansion of the universe.

If an isothermal cylinder does form, one would expect the
hydrodynamic forces that stabilize it to compete with
the gravitational pull of the dark matter that is 
collapsing to form the halo of the galaxy.  This
effect would be expected to retard the flow of the gas
relative to that of the dark matter.  Only the excess gas that cannot
be accomodated into the cylinder structure would be free to
proceed unhindered.  The situation might describe the
filaments of the lingering-gas galaxies.

A measure of the mass per unit length available to
form a filament can be obtained by dividing the total
mass of the turnaround sphere by its proper diameter,
a quantity we will
call the \textit{available mass per unit length},
Figure~\ref{excessgas} shows histograms of the available
mass per unit length for the three galaxy types.  The
histogram for the lingering-gas galaxies peaks at the
mass per unit length value corresponding to a sound speed
squared of one in Figure~\ref{figcorr}, Figure~\ref{figcorrb},
and Figure~\ref{tbtfcssq}.
This value, which marks the end of
reionization, is indicated by the vertical,
dark green, dashed line.   The retreating-gas galaxies peak
to the left and the plunging-gas galaxies to the right.

\cmdbaryloss

The change in baryon fraction between turnaround and
the final collapse sphere for the three galaxy types is
shown in Figure~\ref{baryloss}.  As expected the 
retreating-gas galaxies lose the most baryons and the 
lingering-gas galaxies lose fewer baryons.   
The plunging-gas
galaxies actually gain baryons. 

\cmdturnavail

As expected, available mass per unit length increases
with the total mass of the final galaxy.  This relation is
shown in Figure~\ref{turnavail}.
Since the turnaround radius, and hence the mass within
it as well, is determined by the average radius of the
galaxy's dark matter at turnaround, a simple
hypothesis is that the log of the available
mass per length should be proportional to
the log of the total galaxy mass with a slope
of two-thirds.   The figure, in which each symbol
represents a single galaxy, shows approximate
agreement with
this simple, self-similar picture.

\subsection{Too Big To Fail Galaxies}
\label{sectiontbtf}

Considerable success has been seen in attempts to match observed
galaxies to haloes seen in simulations based upon the $\Lambda$CDM
model (for a review see \citet{somerville15}).  A circular 
velocity for observed
galaxies can be derived from the Doppler broadening of HI lines.
A reasonable relation between this measurement and the
circular velocity of simulated haloes
can be adduced which leads to rough agreement
between the number densities of observed and simulated galaxies
as a function of circular velocity.

However, important discrepancies remain at low halo masses.
The term ``too big to fail'' has been applied to haloes having
a circular velocity between about $25$ 
and $45 \unit{km} \unit{s}^{-1}$
\citep{boylankolchin09,boylankolchin11,boylankolchin12,ferrero12,
garrisonkimmel14,tollerud14,papastergis15,klypin15,papastergis16}.
The frequency of these galaxies in observations is much lower than
what would be predicted from the $\Lambda$CDM model.
The choice of terminology
comes from the paradox that these galaxies are not seen
despite apparently being
massive enough to retain much or all of their gas throughout
reionization.  This situation is in contrast to galaxies 
with lower circular velocities, whose
relative absence from observation is more easily explained
by reionization.

\cmdtbtf
\bold{The ``too big to fail'' concept is based on observations of
present day galaxies, whose detailed histories are uncertain.
However, the simulated galaxies we studied are expected to
include the  mass range of the ``too big to fail'' phenomenon,
and so might be relevant models.}
The scatter plot of Figure~\ref{tbtf} shows the baryon
fraction of the final collapse sphere as a function of the circular 
velocity of its dark matter as estimated from
Equation~\ref{vcircequation}.
\begin{equation}
\label{vcircequation}
V_c = \sqrtsign{\frac{G M_d}{R}}
\end{equation}
where $V_c$ is the circular velocity, $M_d$ the mass of
dark matter, $R$ the radius of the sphere, and $G$ the
gravitational constant.
The approximate circular velocity range of the
``too big to fail'' galaxies is delimited by the vertical, light blue,
dotted lines.  These galaxies have significant baryon deficits
that might make them difficult to observe.    The
green squares represent the galaxies where the gas collapses
as fast or faster than the dark matter, ie. the plunging-gas
galaxies.    The baryon fraction
for these galaxies is close to the cosmic value.  The red
triangles represent the retreating-gas galaxies, 
where the gas fails to
turnaround.  As might be expected these are clustered at
the low end of the circular velocity range and have
the greatest baryon deficits.  The black, filled circles are the
lingering-gas galaxies.  We suggest that the lingering-gas
galaxies are the ``too big
to fail'' galaxies.

We suppose
that the formation of gravitationally bound, isothermal
cylinders inhibits the collapse of gas into haloes
otherwise thought to be massive enough to have
plenty of gas.  These galaxies are then missing from
observations because they form too few stars, that is,
they are ``too big to fail'' but fail because of the
structure of the filaments.

\subsection{An Alternative Dark Matter Model}
\label{sectiondarkmodel}
\boldemp{In this section we consider an alternative filament model
in which the gas is passively trapped in an overwhelming 
dark matter potential determined
by the aggregation properties of dark matter.  
Could such a model be consistent with the filaments we 
have studied?

We do not favor this model because the
filaments are undergoing reionization during the time
in question, presumably by ionizing radiation coming in
from the outside.  
It is not clear theoretically how external ionizing radiation
would affect the aggregation properties of the dark
matter except through its effect on the gas.
Even a collection of the smallest galaxies has
filament gas with the full range of sound speed during
reionization.

To evaluate such a model we computed the bulk velocities of
the dark matter and gas as a function of transverse distance
from the filament axis.  

To improve the accuracy of a velocity
profile, we refined the determination of the position and
orientation of the filament axes.  Anticipating that the gas
velocities would be minimal along the axis, a principal 
component analysis was done using just the gas particles
having a tranverse velocity less 
than $5.38 \unit{km} \unit{s}^{-1}$ \footnote{All velocities
are computed relative to a coordinate system moving with
the average velocity of the gas particles within
311 comoving kiloparsecs of the centre of the galaxy.}.
The orientation of the newly determined axes was close to
that of the original ones,  and considerably greater 
directionality was achieved.  

Figure~\ref{vprof} compares the velocity dispersion
profiles of the dark matter and the gas of filaments from
lingering-gas galaxies whose average sound speed squared
is less than or equal to one in our usual units (see 
Figure~\ref{figcorr}).  These are the filaments we have
found to best match the predictions of the isothermal model we
have put forward in this paper.   Despite the uncertainties
inherent in this computation, it is striking that the gas velocities
are reduced near the axis as would be expected from
hydrodynamic effects.  The dark matter velocities, by contrast,
are greater and are roughly constant out to nearly 
the assumed extent of the filament.  

\cmdvprof

\citet{eisenstein97} have derived a relation between 
the transverse
velocity dispersion  and the mass per unit 
length, $\Upsilon$, 
of a collisionless cylinder at virial equilibrium.
\begin{equation}
\label{velequation}
\Upsilon = \frac{v^{2}}{G}
\end{equation}
where $v$ is the average of the square of the transverse velocity 
and $G$ is the gravitational
constant \citep{eisenstein97}\footnote{This differs from equation
13 of these authors by a factor of 2.  Their equation refers to 
line-of-sight velocity.}.
The relation is similar to that for a gravitationally 
bound, isothermal gas cylinder (equation~\ref{isoequation})
with the bulk velocity substituted for the sound speed.

\cmdtally

Our strategy, to better understand the role of the dark
matter, is to examine the total virial budget, taking
into account both the sound speed of the gas and the bulk
velocities of the dark matter and gas. In an hypothesized
virial equilibrium each source of kinetic energy must
be balanced by a mass per unit length for the filament.
For simplicity, we have taken the mass per unit length of 
the gravitationally bound, isothermal cylinder as that
necessary to balance the sound speed of the gas.  For
the bulk velocities we have taken the mass per unit length for
the collisionless counterpart described above
(Equation~\ref{velequation}).

Figure~\ref{tally} shows that the actual mass per unit
length is consistently too small for both gas and dark matter 
to be present together  in a  virial equilibrium.  This figure
is a histogram for filaments from lingering-gas galaxies
having a sound speed squared
less than or equal to one, using our usual units,  These are the
filaments we have found to match our isothermal cylinder
model the best as shown in Figure~\ref{tbtfcssq}.
The bin on the abcissa is computed by taking the total
mass per unit length of the filament and dividing it by
the mass per unit length required to balance both the
sound speed of the gas and the bulk velocities of the
dark matter and gas.  

These results make it hard to justify a model in which the
gas is part of a dark matter structure at virial equilibrium.
Taking the confines of the filament out to a larger radius
may not solve the problem.  The periphery typically has 
an overdensity of only a few compared to a central overdensity
in the thousands.   Furthermore, the otherwise 
mostly constant velocity of the dark matter increases at large
radius. 

We suggest that only the gas of these filaments 
is close to virial equilibrium.
This supposition fits with the fact the lingering-gas
galaxies are the ones that are the best match to a
gravitationally bound, isothermal gas cylinder.  In these
galaxies
the dark matter collapses into the forming galaxy 
more rapidly than does the gas.
The moving dark matter provides
an environment in which the gravitational constant
appearing in the isothermal gas cylinder equations
is effectively altered because of the potential energy of 
the dark matter. }

\subsection{Number and Alignment of Filaments}
\label{sectionalign}

If the filaments in the simulation correspond
to the rods of the model,
we might expect to see for each galaxy two filaments
protruding from the centre of the galaxy that are roughly
aligned end-to-end.

\cmdnumpeak

Figure~\ref{numpeak} shows that
the number of filaments
identified for each galaxy is generally no more than 
two\footnote{The filaments used for the graphs in this
section are taken from the collection of filaments 
before they are culled by the range test and the
sphere of influence.}.
The six different curves are histograms of the number of
filaments for a galaxy at the six redshifts studied.
In keeping with the model, in the few cases where more
than two were found, only the two most
massive were retained for further analysis.

\cmdpeakalign

When two segments are
present at the same time for a galaxy, the preferred angle
between their directions is close to $180$ degrees,
as if the segments were part of a single, straight rod passing
through the centre of collapse.
This finding dovetails
nicely with the simple paradigm for successive collapse in three
dimensions.

The black, solid line with squares in Figure~\ref{peakalign}
shows a histogram of the angles between the directions of
pairs of segments.  Results for the six redshifts have been
pooled, since the individual results are very similar.  The red,
dotted line with triangles shows the same histogram with
entries weighted by the reciprocal of the sine of the angle.
This latter plot is suited to a situation in which the two
filaments are assumed to come together in three dimensions
rather than forming within a predetermined plane.

In this paper, for clarity,
we have called each rod protruding from the centre of the
galaxy a filament
even though it is attractive theoretically to envision
the collapse of a
single filament passing through the centre of the galaxy.

\section{Discussion}
\label{discussion}

The theme of this paper is that the gas in the intergalactic filaments
during the first gigayear
can be best understood in terms
of simple hydrodynamic principles.
In contrast, the dark matter
can be understood in terms of the spherically symmetric
collapse of an overdensity in an expanding universe.

\boldemp{The filaments undergoing reionization fit the
isothermal model best.  These filaments tend to
have dark matter
spread roughly evenly over the entire length.  
Isothermal cylinders may also
be present at later times, but be harder to demonstrate
because large dark matter clumps may negate
the assumption of uniform mixing.}

It is important to emphasize that the filaments we see
should be distinguished
from the generally larger ones others study
at lower redshifts, which often contain within them
multiple galaxies.  Most papers on the cosmic
web have dealt with these massive, later structures.

\citet{eisenstein97} has described a method for determining the
dynamical mass per unit length of a massive filament containing
hundreds of galaxies such as might be observed in redshift
surveys at low redshift.  The method can be used to estimate
the mass to light ratios of these filaments.  The method is impressive
and has theoretical antecedents in common with our model.
However, the method is not easily applicable to the very
small filaments associated with the galaxies of the first
gigayear.

  \citet{danovich}  have reported that galaxies
tend to be at the centre of three filaments
contrary to
our findings.  We also often see more than two
filaments extending from a single point.  However, we
believe that these are likely to be the
result of later collisions among filaments.
Most of the $1{,}200$ largest galaxies can be
traced back to an early stage
when no more than two roughly end-to-end filaments
are present.

To study the numbers and properties of filaments, 
it is necessary to
identify them in an algorithmic fashion free of subjective
bias.  We find that visual inspection
can be deceptive.  A plane
viewed edge on or the intersection of two planes can appear
as a spurious filament.  In addition, some
of the filament segments are difficult to pick out from the
background by eye.

We believe that our criteria for selection of filament
candidates are good, and that it is not unreasonable
to discard additional faint filaments that may be seen by eye.

It should also be appreciated that, at early times,
the centre of a future galaxy connecting two end-to-end
filaments may not be apparent upon casual
inspection.  Rather it may appear to the eye
simply as a point
on a single continuous filament.
This situation could lead to an apparent
preference for
more than two filaments emanating from a
single galaxy.

We argue that, during the reionization process, the gas
assumes a structure that can be understood in terms of a
gravitationally bound, isothermal
cylinder at equilibrium at about $10^{4}\unit{K}$.
We find that this temperature, close to the Lyman alpha cooling
floor, is tightly maintained during reionization.  This
temperature regulation may be important in the
stability of the cylinders during this period.

The change in sound speed during
reionization can be seen reflected in a change in
the mass per unit length of the cylinder (Figure~\ref{figcorr}).
We consider this finding together with the proportionality
constant to be strong evidence for gravitationally
bound, isothermal cylinders.  

\boldemp{To further justify our emphasis on the importance of the
gas component, we explored the virial consequences of
a more conventional model in which the structure is
primarily determined by the aggregation properties of
dark matter.  There appeared to be too little mass per
unit length for such a model.}

It is important to note
that the simulation follows in detail the ionization of the gas by
radiation from the actual sites of stellar formation.
This is in contrast
to many simulations that merely impose a uniform
radiation field and thus might miss such structural
nuances.  \boldemp{For such simulations 
we would not expect 
dramatic changes in the physics of the gas in the
filaments because there is very little stellar material
in the filaments.  However, the time course of
reionization would probably be altered with the result that
there would be fewer filaments with intermediate
ionization stages.  Demonstration of isothermal
cylinders might then be more difficult.}

We find remarkable uniformity in the behaviour of
the dark matter of the $1{,}200$ largest galaxies
in the simulation.  These galaxies range in
total mass, including dark matter, gas, and
stars, over two orders of magnitude.  The dark
matter collapse follows closely the spherically
symmetric cycloid model at least through
turnaround.  The overdensity at turnaround agrees
roughly with theoretical expectation.  The matter
collapses initially into a rough plane and then  into
a one dimensional structure within the plane,
resulting in up to two filaments protruding from the
centre of the galaxy oriented end to end.  At any time,
approximately three quarters of the galaxies have at least
one filament emerging from it.

An outstanding problem in cosmology is the
discrepancy between the observed number of satellite
galaxies of the Milky Way and the predicted number
from simulations using dark matter only \citep{kauffmann93}.
Our results suggest a mechanism that suppresses the
entry of gas into dark matter haloes.  This
phenomenon would be in contrast to the expulsion of
gas from haloes by heating and photoionization.

One might imagine that filaments would
facilitate the entry of gas into the halo by 
restricting its angular
momentum.  However, our results argue that, on balance, the
gas that can be accomodated into an isothermal cylinder
is retarded in its motion. 

\boldemp{A simulation with greater resolution would be 
desirable to confirm and expand upon these
results.  Higher resolution might 
reveal structures 
important in the formation of galaxies
smaller 
than the ones studied here.  It would be 
interesting to see if there is a minimum galaxy
size that can form according to our filament
model.  Would very small galaxies form very
early when the temperature is substantially
lower than the Lyman alpha cooling floor?

A higher resolution would also help to establish
whether the dark matter really is as clumpy
compared to the gas as it appears.  
Sometimes increasing resolution can expose
artifacts arising from the discreteness of the
simulation elements.   We would not expect
a higher resolution to reveal the gas in the 
filaments as fragments
rather than continuous.  This is because, when discreteness is
important, it is usually the lower resolution simulation
that is fragmented rather than the higher one.  For a
discussion of discreteness effects see \citet{power16}.}

\boldemp{Reionization is regarded as a watershed event in
the history of the universe.  A galaxy that formed before
reionization could retain an ancient population of stars that
could not have formed afterward if the galaxy were too small.  
\citet{brown14} have argued
from an analysis of individual stars that some ultra-faint dwarf
satellites of the Milky Way are indeed such galaxies that have 
survived to the present day.  The galaxies in our study all began
their formation prior to reionization.  We do not know their
likely fate, but it is plausible that they could
share common features with some small galaxies observed at redshift zero.
It is thus not unreasonable to compare our galaxies with galaxy 
populations observed today that exhibit the ``too big to fail''
effect.}

A comprehensive consideration of the ``too big to fail'' phenomenon
is beyond the scope of this paper.  We note, 
however, that our model differs
from most of the other ones in focussing on the transport of
gas into the galaxy, rather than on baryonic effects and
star formation within the galaxy (see for example,
\citet{governato12,zolotov12,brooks13,veraciro13,arraki14,brooks14,
madau14b,jiang15,pawlowski15,agertz16,read16,wetzel16,zhu16}.

A detailed analysis of the motion of gas into
galaxies is beyond the scope of this paper.
Our model is consistent with the ``cold mode''
of galaxy accretion.
We see no evidence for a hot gas stage in the
formation of our galaxies.  We have argued from rates
of collapse of the dark matter and gas that the gas
moves into the galaxy along a single, linear filament.

An important open question involves the timing
and mechanism of the reionization of the universe.
To evaluate the role of ionizing radiation from
stars one must know the fraction of stellar radiation
that escapes from its point of origin.  It is clear that the
gas surrounding the centre of collapse of a galaxy is not
spherically symmetric.  Our results may provide a more
realistic starting point for such calculations.

Finally, our model suggests the central presence of gas at 
a very early time in the formation of the galaxy halo.
Perhaps this gas contributes to the formation of
a ``cored'' halo profile as
opposed to the ``cuspy'' one suggested from simulations
containing only dark matter \citep{gilmore07,evans09,
deblok10,strigari10,amorisco12,martinez15}.

\section{Summary}
\label{sectionsummary}

We propose a model for the development of intergalactic
filaments during the first gigayear of the universe.
In this model the mass per unit length and structure
of the filament is
determined, not by the potential well of the enclosing
dark matter, but by the hydrodynamic properties of the gas.

The model is described in the context of galaxy formation.
Up to two extended filaments may protrude
from the centre of the collapsing galaxy.  They tend to be
oriented end-to-end as if they comprised a single structure.
The total mass per unit length of a filament segment
is proportional to the square of the sound speed of the
gas with a
proportionality constant equal to that predicted for
a gravitationally bound,
isothermal cylinder.  The sound speed of a gas filament
depends only upon
its temperature and ionization state 
and not upon the density.
These structures generally contain gas and dark
matter in roughly the cosmic ratio.  The dark matter
contributes to the gravitational field in proportion to its
abundance as if it were uniformly mixed with the gas.

The dark matter of each galaxy collapses according to
the simplest model for a spherically symmetric overdensity in
an expanding universe.  This cycloid profile is followed
until some time after turnaround.  The overdensity of the
material at turnaround agrees with theoretical
predictions.

After reionization
the average gas fraction of a filament
segment may increase as the collapse of dark matter progresses.

Overall the dark matter collapses to roughly the same
extent for each galaxy.  However, the gas collapse varies.
Three types of galaxies are distinguished.
In the ``plunging gas'' galaxies, the gas collapses
as fast or faster than the dark matter.  These galaxies appear
to collapse from overdensities having far too much matter
to form a gravitationally bound, isothermal filament under the
ambient thermal conditions.  In the ``retreating gas'' galaxies,
the gas fails to move toward the halo and may expand with the
universe.  These galaxies appear to
collapse from overdensities having too little matter.
Finally, in the ``lingering gas'' galaxies the gas collapses but more
slowly than does the dark matter.  These appear to 
have available masses
in a suitable range to form gravitationally bound, isothermal
cylinders.  Presumably, the gas that can be accomodated in
the isothermal structure is retarded
in its motion, whereas excess gas can can proceed unhindered.

For the lingering gas galaxies
the different overall rates of migration of the dark
matter and gas predict the final baryon fraction of the
galaxy under a simple assumption of uniform, filamentary
accretion, but are inconsistent with spherical
accretion.

The model works for a galaxy mass range that includes
that of the so-called ``too big to fail'' galaxies and may explain
the peculiar absence of these galaxies from
observational surveys.

The model may provide a mathematical framework for
understanding a variety of open questions about
structure formation in the early universe.  The sizes
of the structures in this model suggest a minimum
simulation resolution that may be necessary to recreate
the effects seen here.  In addition, the model may
contribute to our intuition
about the roles of dark matter and
gas in structure formation.

\section*{Acknowledgments}
We are grateful to Nickolay Y. Gnedin for
providing us with
the output of his simulation
along with his visualization software IFRIT.

\bibliographystyle{mn2e}

\bibliography{harford}

\appendix

\section{Algorithm For Filament Identification}
\label{filamentid}

Each filament is selected in the context of a forming
galaxy.  Figure~\ref{geom} shows the geometry
for this selection for an example galaxy.
The selection takes place in a comoving coordinate system
whose centre is the centre of mass of the collapsing 
dark matter particles of the galaxy.  Unlike the gas
particles, the individual dark matter particles of a 
galaxy can be followed throughout the simulation.
The XY plane is set to the plane of 
the gas as determined by a principal component
analysis of the gas particles within a 
sphere of radius $266$ kpc.
The large black rectangle in the
Figure~\ref{geom} shows this XY plane.  Annuli above (red) and
below (green) the plane
determine a volume that is divided into equal 
bins by the light blue
rectangles.  The intersection of the bins
with the plane are shown by the black annulus.
The annulus is defined
by two radii of 88.8 and 133.2\unit{kpc}.

\cmdgeom
\cmdfilselect

A bin having
more than twice the average amount of gas per 
bin is considered to
contain a filament.  In cases where neighboring 
bins of the annulus both meet this criterion, 
the gas in both bins is merged
and analyzed together.
The filament identification process is completely defined by
this algorithm.  There are no subjective elements.

The algorithm identifies
visually clear segments that are usually part of
longer filaments extending radially
from the centre of collapse of the galaxy.   Also identified,
and therefore
included in our analysis, are structures that may not be
readily identified or interpreted by eye.

These filament segments are used in 
Section~\ref{sectionalign} to
compare the relative orientations
of the filaments with the prediction of the model.
However,
for the analyses in the remaining sections, the set of segments
is further culled using a ``range test''.  In this test, the
filament segment is divided into three parts.  
To pass the test,
the mass of gas in each part can differ from the 
average of the three parts
by no more than 20\%.  This test ensures that
the identified
structure grossly resembles a filament.  
This is important because
we do not allow any subjective inspection of a segment to
influence whether it is included.

In addition, we require that the sphere of influence of the
galaxy cover at least half of the filament segment in question.
This is done to ensure that nearby galaxies do not
result in the selection of the same filament.
The sphere of influence is considered to be a
sphere with radius equal to the turnaround radius in comoving
coordinates centred at the centre of the galaxy at the
time in question.
  
A total of $2300$ filament segments selected from the six redshifts
considered survived all tests.
 
For structural analysis
the orientation of each segment
is determined more precisely using a principal
component analysis.

\end{document}